\documentclass[preprint,12pt]{elsarticle}

\usepackage{amssymb}
\usepackage{amsmath}
\usepackage[ruled,vlined]{algorithm2e}
\SetKw{KwTo}{to}
\SetKw{KwStep}{step}
\usepackage{booktabs}
\usepackage{multirow}
\usepackage{array}
\usepackage[margin=1in]{geometry}
\usepackage{booktabs}

\journal{}

\begin{document}

\begin{frontmatter}

\title{ELAS3D-Xtal: An OpenMP-accelerated crystal elasticity solver with automated experiment-driven microstructure generation}

\author[label1]{Juyoung Jeong}

\author[label1]{Veera Sundararaghavan\corref{cor1}}
\ead{veeras@umich.edu}

\cortext[cor1]{Corresponding author}

\affiliation[label1]{organization={Department of Aerospace Engineering},
            addressline={University of Michigan}, 
            city={Ann Arbor},
            postcode={48109}, 
            state={MI},
            country={USA}}

\begin{abstract}
This paper introduces \texttt{ELAS3D-Xtal}, a high-performance Fortran/OpenMP upgrade of the NIST ELAS3D voxel-based finite element solver for computing 3D elastic fields in polycrystals with defects. The code supports crystal anisotropy by precomputing rotated stiffness tensors from user-specified orientations and solves the equilibrium problem with a matrix-free, OpenMP-parallel preconditioned conjugate-gradient (PCG) method using a point-block Jacobi preconditioner. On a single shared-memory multicore PC, OpenMP threading accelerates the baseline CG solver by $\sim 10\times$, while the block-preconditioned CG solver achieves $53$--$61\times$ speedup relative to the serial CG baseline for meshes from $100^3$ to $500^3$ voxels (scaling to domains up to $800^3$ voxels). Accuracy is validated against the analytical Eshelby inclusion solution. \texttt{ELAS3D-Xtal} also integrates microstructure construction, including statistically calibrated polycrystal generation via spatial filtering and parallel voxel-to-grain assignment, direct pore insertion from XCT centroid/radius data, and texture assignment. Full-field phase, orientation, and stress outputs are written in HDF5 to enable scalable post-processing and defect-mechanics workflows. Applications are demonstrated for (i) anisotropy-controlled defect-scale stress fields and (ii) LPBF SS316L microstructures with gas, lack-of-fusion, and keyhole pore morphologies.
\end{abstract}


\end{frontmatter}

\section{Introduction}
\label{sec1}
Simulating the elastic response of realistic three-dimensional (3D) polycrystals and defect-containing microstructures is essential for benchmarking, process modeling, and materials design. Accurate simulation of 3D elastic fields is particularly essential for predicting local stress concentrations and fatigue life in additively manufactured (AM) metals. In additive processing methods, pores resulting from a lack of particle fusion, keyhole defects, and gas entrapment, which are typically tens of micrometers in size, act as critical stress risers. Furthermore, these defects are embedded within anisotropic grain matrices of comparable scale (tens of microns), requiring high-resolution discretizations (e.g., 5–10 $\mu m$ voxels) to accurately resolve stresses in large-scale domains ($10^6$-$10^9$ voxels). The computational difficulty in the accurate prediction of fatigue life of as--built additively manufactured specimens limits the widespread usage of such parts in load-bearing structures. 
The main challenge is to achieve scale and physical fidelity in the presence of strong material heterogeneity. General-purpose finite element software such as Abaqus~\cite{abaqus2011abaqus} or COMSOL~\cite{multiphysics1998introduction} offers mature constitutive models but imposes substantial meshing and memory overhead due to the lack of native voxel-based crystal elasticity and automated microstructure handling. Conversely, FFT-based spectral solvers~\cite{lucarini2022fft} can be efficient for periodic heterogeneous media, but porous AM microstructures introduce severe stiffness contrasts that can lead to oscillatory artifacts near void--solid interfaces. The original NIST ELAS3D code~\cite{garboczi1998finite}, based on a finite element framework, provided an effective scheme for such problems, but its scalability and convergence were limited by single-threaded execution and basic iterative solvers. A parallelized version~\cite{bohn2003user} was introduced based on the Message Passing Interface (MPI)~\cite{gropp1999using} domain decomposition. MPI remains the standard for extreme-scale simulations ($>10^{10}$ voxels)  across distributed supercomputing clusters in many fields, including crystal plasticity~\cite{yaghoobi2019prisms,jeong2022physics, jeong2023crystal}, molecular dynamics~\cite{thompson2022lammps}, astrophysics~\cite{xia2018mpi}, plasma simulation~\cite{dudson2009bout++}, and computational fluid dynamics~\cite{jasak2007openfoam}. However, for large problems on a single machine with many cores and large memory, MPI-style domain decomposition can introduce unnecessary overhead, such as ghost layers and subdomain synchronization, even though all data reside in shared memory. Furthermore, legacy solvers lack the advanced preconditioning required to handle ill-conditioned stiffness matrices arising from the infinite contrast between stiff metal grains and voids.
To address these limitations, the upgraded ELAS3D-Xtal code presented here builds on the NIST ELAS3D framework~\cite{garboczi1998finite}, introducing high-performance parallelization, advanced preconditioning, full support for crystal anisotropy and experimental microstructure data, and modern output capabilities. The key additions include: (i) a standalone Fortran/OpenMP~\cite{dagum1998openmp} implementation that utilizes shared-memory threading; (ii) voxelwise crystal anisotropy with orientation handling using Rodrigues rotations; (iii) a parallel preconditioned conjugate gradient (PCG) solver with block Jacobi preconditioning; (iv) modern HDF5-based output for scalable post-processing and simplified data exchange; and (v) full field stress outputs.
Experimentally, microstructure and defects can be measured in additively manufactured specimens using diffraction microscopy, tomography, and in-situ monitoring methods. However, leveraging data from these modalities to build microstructural simulation volumes for input into elasticity codes remains a challenge. Many existing tools struggle to account for the combined effects of grain morphology, defects, and texture. Another key advantage of the current code is its native microstructure construction method, which generates polycrystals with assigned grain size and aspect ratio, while also assigning correct crystallographic texture and pore distributions. Most current solvers lack integrated microstructure generators; tools such as Neper~\cite{quey2011large,quey2018optimal} and DREAM.3D~\cite{groeber2014dream} provide statistical polycrystals, but require format conversion and separate pore embedding from X-ray computed tomography (XCT) data, without direct solver coupling. ELAS3D-Xtal embeds a native microstructure generator using a fast parallel ellipsoidal Voronoi tessellation method. 
Section~\ref{sec2} presents the solver upgrades for computing equilibrium displacement and stress fields in large voxel-based domains, along with the microstructure generation algorithms that enable defect-resolved simulations. Section~\ref{sec3} validates the framework using the 3D Eshelby inclusion benchmark and mesh sensitivity/scaling studies, and then demonstrates defect-scale stress fields for (i) controlled elastic-anisotropy variations and (ii) LPBF SS316L microstructures with representative pore morphologies. Section~\ref{sec4} discusses the implications and practical considerations for applying the framework to AM defect mechanics. Lastly, Section~\ref{sec5} summarizes the main findings and describes future work.

\section{Methods}
\label{sec2}
\subsection{Three-dimensional finite element elastic solver}
\label{sec2:1}
The elastic response of the synthetic microstructure is computed using a custom high-performance finite element solver, \texttt{ELAS3D-Xtal}. This solver builds upon the foundational voxel-based framework of NIST ELAS3D~\cite{garboczi1998finite} but has been re-architected for modern shared-memory parallel systems.
\subsubsection{Discretization and constitutive formulation}
\label{sec2:1:1}
The computational domain is discretized as a regular three-dimensional grid containing $N = N_x \times N_y \times N_z$ voxels. Each voxel is treated as a trilinear hexahedral finite element following the foundational methodology of the NIST ELAS3D solver~\cite{garboczi1998finite}. To simulate a representative volume element (RVE) embedded in an infinite medium, periodic boundary conditions (PBCs) are enforced via a neighbor-lookup table $\texttt{ib}(m, n)$. This table maps every node $m$ and the local neighbor index $n \in [1, 27]$ to a global memory index using grid coordinates such that $\texttt{ib}(m,n) \mapsto \text{index}(i',j',k')$:
\begin{equation}
\begin{aligned}
i' &= 1 + \operatorname{mod}\!\bigl(i+\Delta i_n-1,\; N_x\bigr),\\
j' &= 1 + \operatorname{mod}\!\bigl(j+\Delta j_n-1,\; N_y\bigr),\\
k' &= 1 + \operatorname{mod}\!\bigl(k+\Delta k_n-1,\; N_z\bigr).
\end{aligned}
\label{eqn:pbc_indices}
\end{equation}
where, $(i, j, k)$ represent the integer grid coordinates of the central node $m$, and $(\Delta i_n, \Delta j_n, \Delta k_n)$ denote the relative spatial offsets to the surrounding nodes. The resulting terms $i', j', k'$ describe the voxel indices of the neighbor $n$ after correcting for periodic wrapping at the domain boundaries.

A significant re-architecture of the original solver involves the handling of constitutive laws. Although the legacy code primarily utilized scalar storage for isotropic phases, the updated framework assigns a full stiffness tensor to every phase or grain to enable crystal elasticity modeling. For isotropic phases, such as the base matrix in simple composites, the stiffness tensor $\mathbf{C}_m$ is constructed analytically using the Lamé constants $\lambda$ and $\mu$, derived from the Young's modulus $E$ and Poisson's ratio $\nu$. For each phase $p$ (grain or pore), an elastic stiffness tensor $\mathbb{C}_p$ is assigned. Isotropic elasticity uses Lamé constants:
\begin{equation}
\lambda_p = \frac{E_p \nu_p}{(1 + \nu_p)(1 - 2\nu_p)}, \qquad
\mu_p = \frac{E_p}{2(1 + \nu_p)}
\label{eqn:2}
\end{equation}
with stiffness tensor components
\begin{equation}
C_{ijkl}^{(p)} = \lambda_p \delta_{ij} \delta_{kl} + \mu_p (\delta_{ik}\delta_{jl} + \delta_{il}\delta_{jk}).
\label{eqn:3}
\end{equation}

For polycrystalline grains, physical fidelity is maintained by assigning full anisotropic stiffness tensors. The local crystal stiffness, $\mathbb{C}^{\text{local}}$, is rotated into the global coordinate system using the specific orientation vector $\boldsymbol{\rho}_p$ assigned to each grain $p$. We implemented an explicit tensor rotation subroutine \texttt{rotatestiffnessby} (Algorithm~\ref{alg:rotatestiffnessby}) that transforms the stiffness from the local (crystal) frame to the global (sample) frame. Transforming the fourth-rank stiffness tensor in Voigt notation requires the construction of a Bond stress transformation matrix, denoted here as $\mathbf{Q}_p$. Given a grain orientation parameterized by a Rodrigues vector $\boldsymbol{\rho}_p$, the solver first computes the standard $3 \times 3$ rotation matrix $\mathbf{R}_p$. The $6 \times 6$ transformation matrix $\mathbf{Q}_p$ is then assembled element-wise from the components of $\mathbf{R}_p$, explicitly mapping the trigonometric products for normal and shear stress interactions. The global stiffness tensor for grain $p$ is thus computed as:
\begin{equation}
\mathbb{C}_p^{\text{global}} = \mathbf{Q}_p \mathbb{C}^{\text{local}} \mathbf{Q}_p^T
\label{eqn:anisotropic_rotation}
\end{equation}
This operation is pre-computed for each unique grain orientation during the initialization phase to minimize computational overhead during the iterative solution. Pore phases are modeled by assigning zero stiffness values to simulate voids.

\begin{algorithm}
\caption{\texttt{rotatestiffnessby}: Pre-compute Rotated Elastic Tensors}
\label{alg:rotatestiffnessby}
\small
\SetAlgoLined
\DontPrintSemicolon

\KwIn{Data structure \texttt{e3d} (containing orientation vectors $r$).}
\KwOut{Populated \texttt{e3d\%rotatedstiffness} (Global $6\times6$ matrices).}

\BlankLine
\tcp{1. Define Local Crystal Stiffness (Cubic)}
$C_{local} \in \mathbb{R}^{6\times6} \leftarrow 0$\;
Fill diagonal elements with $C_{11}, C_{44}$\;
Fill off-diagonal interaction terms with $C_{12}$\;

\BlankLine
\tcp{2. Loop Over Phases}
\For{$g = 1$ \KwTo $nphase-1$}{
    
    \BlankLine
    \tcp{A. Compute Orientation Matrix ($3\times3$)}
    $r \leftarrow$ Orientation vector for phase $g$\;
    $R_{3\times3} \leftarrow \textsc{RodriguesFormula}(r)$ 
    
    \BlankLine
    \tcp{B. Construct Bond Transformation Matrix ($6\times6$)}
    Construct $M_{bond} \in \mathbb{R}^{6\times6}$ using elements of $R_{3\times3}$\;
    
    \BlankLine
    \tcp{C. Rotate Stiffness Tensor}
    $C_{rot} \leftarrow M_{bond}^T \cdot C_{local} \cdot M_{bond}$\;
    
    \BlankLine
    \tcp{D. Store in Global Array}
    Flatten $C_{rot}$ ($6\times6$) and store into \texttt{e3d\%rotatedstiffness} at index $g$\;
}

\BlankLine
\Return{Updated $\mathtt{e3d\%}$rotatedstiffness}\;
\end{algorithm}

\subsubsection{Parallel preconditioned conjugate gradient solver}
\label{sec2:1:2}
The equilibrium displacement field $\mathbf{u}$ is obtained by enforcing the principle of minimum potential energy. The problem is prescribed by an imposed macroscopic strain tensor, $\boldsymbol{\mathcal{E}}$, which defines the global boundary condition. The initial displacement at node $m$, denoted $\mathbf{u}_m^0$, is taken as an affine mapping of the node’s reference coordinates $\mathbf{X}_m$:
\begin{equation}
\mathbf{u}^0_m = \boldsymbol{\mathcal{E}} \cdot \mathbf{X}_m
\quad \Longleftrightarrow \quad
(u_i)_m = \sum_{j=1}^{3} \mathcal{E}_{ij}\,(X_j)_m .
\label{eqn:affine_strain}
\end{equation}
This formulation ensures that the initial guess is a vector consistent with the applied deformation gradient, serving as the starting point for the relaxation process.

The total potential energy $\Pi(\mathbf{u})$ of the system is defined as the sum of the internal elastic strain energy stored in all elements and the potential energy of external forces. In the continuum limit, this is expressed by: 
\begin{equation}
\Pi(\mathbf{u}) = \int_{\Omega} \frac{1}{2} \boldsymbol{\epsilon} : \mathbb{C} : \boldsymbol{\epsilon} d\Omega - \int_{\Gamma} \mathbf{t} \cdot \mathbf{u}  d\Gamma.
\label{eqn:continuum_energy}
\end{equation}

In the present voxel-based finite element framework, the work contributions from the macroscopic strain and boundary constraints are assembled into a constant term $C$ and a linear force vector $\mathbf{b}$. Upon discretization into trilinear hexahedral finite elements, the total energy becomes the summation of the discrete energy functional from individual voxels:
\begin{equation}
\Pi(\mathbf{u}) =C+\sum_{e=1}^{N_{\text{elem}}} \left( \frac{1}{2} \mathbf{u}_e^{T}\mathbf{K}_e \mathbf{u}_e \right) + \mathbf{b}^{T}\mathbf{u}
\label{eqn:discrete_energy}
\end{equation}
where $\mathbf{u}_e$ denotes the vector of nodal displacements for element $e$, $\mathbf{b}$ represents the residual force vector resulting from the applied boundary constraints, and $\mathbf{K}_e$ represents the local element stiffness matrix derived from the constitutive laws in Section~\ref{sec2:1:1}. Assembling these contributions leads to the global quadratic minimization problem:
\begin{equation}
\min_{\mathbf{u}} \Pi(\mathbf{u}) = \min_{\mathbf{u}} \left(C + \frac{1}{2}\mathbf{u}^{T}\mathbb{A}\mathbf{u} + \mathbf{b}^{T}\mathbf{u}\right).
\label{eqn:quadratic_form}
\end{equation}
where $\mathbb{A}$ is the global Hessian (stiffness) matrix. Differentiating this functional yields the global force gradient (variable \texttt{gb} in the code): $\nabla\Pi(\mathbf{u}) = \mathbb{A}\mathbf{u} + \mathbf{b}$. The equilibrium displacement field is the solution to the linear system $\mathbb{A}\mathbf{u} + \mathbf{b} = \mathbf{0}$.

For microstructures with more than $10^8$ degrees of freedom, explicitly assembling the global stiffness matrix $\mathbb{A}$ is memory-prohibitive. Consequently, the core solution procedure was re-architected into a matrix-free parallel PCG framework, implemented in the subroutine \texttt{dembx\_OpenMP} (Algorithm~\ref{alg:dembx_OpenMP}). This subroutine introduces two specific innovations to address the computational challenges posed by high-contrast polycrystalline anisotropy: (1) point-block Jacobi preconditioner and (2) cache-blocked matrix-free product.

Standard legacy solvers often employ a scalar Jacobi preconditioner, i.e., component-wise scaling by $1/\mathbb{A}_{ii}$. In anisotropic crystals, however, the three displacement components at a node $(u_x,u_y,u_z)$ are strongly coupled through off-diagonal terms in the local nodal stiffness, so scalar scaling can be ineffective. We therefore employ a block-diagonal (block-Jacobi) preconditioner $\mathbf{M}$ that treats the three degrees of freedom at each node as a dense coupled $3\times 3$ block~\cite{saad2003iterative, axelsson1996iterative}. During the initialization phase of Algorithm~\ref{alg:dembx_OpenMP}, the diagonal $3\times 3$ block $\mathbb{A}_{mm}$ associated with node $m$ is assembled by summing the corresponding $3\times 3$ element sub-blocks $(\mathbf{K}_e)_{mm}$ over all neighboring elements $e\in\mathcal{N}(m)$ (eight voxels for an interior node), and the resulting block is inverted using a cofactor-based closed-form $3\times 3$ inverse:
\begin{equation}
\mathbf{M}_m^{-1} = \left(\mathbb{A}_{mm}\right)^{-1}
= \left( \sum_{e \in \mathcal{N}(m)} (\mathbf{K}_e)_{mm} \right)^{-1}.
\label{eqn:block_precond}
\end{equation}
This precomputation improves the local conditioning of the system and helps mitigate convergence stagnation in regions with strong stiffness contrast, such as at grain boundaries or near voids.

The dominant computational cost is the matrix--vector product of the directional stiffness, $\mathbf{Ah}=\mathbb{A}\mathbf{h}$, which is evaluated matrix-free using the 27-point stencil. Since the kernel is largely memory-bandwidth bound, which means repeatedly loading neighboring values of the virtual displacement field $\mathbf{h}$ and the associated lookup/stencil data, we introduce manual loop blocking, including outer loops $L1$ and $L2$ in Algorithm~\ref{alg:dembx_OpenMP}. Specifically, the node index is processed in contiguous chunks of size \texttt{block\_size}, which improves cache locality and increases the likelihood that the relevant portions of  $\mathbf{h}$ and the accumulated $\mathbf{Ah}$ remain in the L1/L2 cache during the stencil evaluation. Compared with an unblocked traversal, the blocked traversal strategy reduces cache-miss overhead and improves parallel efficiency on multicore CPUs.

The performance of this approach depends on the choice of \texttt{block\_size}, which determines the inner kernel's working set size. For double-precision data, 8 bytes per scalar, storing $\mathbf{h}$ and $\mathbf{Ah}$ for $N_{\text{block}}$ nodes requires approximately $6N_{\text{block}}\times 8 \approx 48N_{\text{block}}$ bytes, not including auxiliary arrays (e.g., neighbor indices and stencil coefficients). In practice, \texttt{block\_size} is tuned so that the working set occupies roughly 50--70\% of the available per-core L1/L2 cache. On modern Intel and AMD processors, this typically corresponds to $N_{\text{block}}\approx 128$--$512$. Blocks that are too large can trigger cache eviction and reduce throughput, whereas blocks that are too small increase loop overhead and diminish the benefits of vectorization.

\begin{algorithm}
\caption{\texttt{dembx\_OpenMP}: Parallel Block-Jacobi PCG Solver}
\label{alg:dembx_OpenMP}
\small
\SetAlgoLined
\DontPrintSemicolon

\KwIn{State \texttt{e3d} ($u, gb, h \in \mathbb{R}^{ns\times 3}$); Stencil: $ib$; Map: $pix$; Coeffs: $dk$; Ref. norm: $gg_0$; Globals: $ldemb, tol, block\_size$}
\KwOut{Updated \texttt{e3d} fields; Final stats $Lstep, gg$.}

\BlankLine
\tcp{1. Build Point-Block Jacobi Preconditioner}
Allocate $u,gb,h,Ah,z \in\mathbb{R}^{ns\times 3}$; \quad Allocate local $M^{-1} \in \mathbb{R}^{ns \times 3 \times 3}$\;
$u\leftarrow e3d\%u$; \quad $gb\leftarrow e3d\%gb$; \quad $h\leftarrow e3d\%h$\;

\BlankLine
\ForPar{$m=1$ \KwTo $ns$}{
  $ibl \leftarrow e3d\%ib(m,:)$; \quad $pl \leftarrow e3d\%pix(ibl)$\;

  \tcp{Assemble local diagonal $3\times 3$ block $M_m$}
  \For{$i=1$ \KwTo $3$}{
    \For{$j=1$ \KwTo $3$}{
      $M_m(i,j) \leftarrow$ sum of contributions from $dk(\cdot)$ via $pl$\;
    }
  }
  $M^{-1}(m,:,:) \leftarrow \textsc{Invert3x3}(M_m)$\;
  $z(m,:) \leftarrow M^{-1}(m,:,:) \cdot gb(m,:)$\;
}

\BlankLine
\tcp{2. Parallel PCG Iteration Loop}
$Lstep \leftarrow 0$; \quad $h \leftarrow z$\;
\For{$iter=1$ \KwTo $ldemb$}{
    $Lstep \leftarrow Lstep + 1$; \quad $Ah \leftarrow 0$\;
    
    \tcp{A. Directional Stiffness (Cache-Blocked OpenMP)}
    \For{$L1=1$ \KwTo $ns$ \KwStep $block\_size$}{
        $L2 \leftarrow \min(L1+block\_size-1, ns)$\;
        \ForPar{$m=L1$ \KwTo $L2$}{
            \For{$j=1$ \KwTo $3$}{
                $Ah(m,j) \leftarrow \sum_{q=1}^{27}\sum_{n=1}^{3} K_{m,q}(j,n;pl)\, h(ibl[q],n)$\;
            }
        }
    }

    \tcp{B. Step Size $\lambda$}
    $hAh \leftarrow \sum_{m,j} h(m,j)\,Ah(m,j)$; \quad \lIf{$|hAh| < 10^{-14}$}{\textbf{break}}
    $gg \leftarrow \sum_{m,j} z(m,j)\,gb(m,j)$; \quad $\lambda \leftarrow gg / hAh$\;
    
    \tcp{C. Update Solution \& Residual}
    $u \leftarrow u - \lambda h$; \quad $gb \leftarrow gb - \lambda Ah$\;

    \tcp{D. Precondition \& Convergence}
    \lForPar{$m=1$ \KwTo $ns$}{$z(m,:) \leftarrow M^{-1}(m,:,:) \cdot gb(m,:)$}

    $gg_{\text{last}} \leftarrow gg$\;
    $gg \leftarrow \sum_{m,j} z(m,j)\,gb(m,j)$; \quad    \lIf{$\sqrt{gg/gg_0} < tol$}{\textbf{break}}
    
    \tcp{E. Update Direction (Fletcher--Reeves)}
    $\gamma \leftarrow gg/gg_{\text{last}}$; \quad $h \leftarrow z + \gamma h$\;
}
\BlankLine
\Return{Updated $\mathtt{e3d}$ structures}\;
\end{algorithm}

\subsubsection{Implementation, workflow, and performance}
\label{sec2:1:3}
The \texttt{ELAS3D-Xtal} solver is implemented as a high-performance modular framework written in Fortran 90, designed for shared-memory parallel execution via OpenMP. Unlike the legacy serial implementation, the updated architecture leverages guided memory allocation and HDF5 I/O libraries to handle microstructure datasets that exceed hundreds of millions of voxels. The code is organized into distinct computational modules that separate data initialization, constitutive mapping, iterative solution, and post-processing. A breakdown of these principal modules and their specific functions is provided in Table~\ref{tab:solver_modules}.

\begin{table}[ht]
\centering
\caption{Principal computational modules in ELAS3D-Xtal}
\label{tab:solver_modules}
\begin{tabular}{@{}ll@{}}
\toprule
\textbf{Module} & \textbf{Function} \\
\midrule
\texttt{femat\_OpenMP} & Stiffness matrix assembly, boundary term setup \\
\texttt{energy\_OpenMP} & Total energy and gradient computation \\
\texttt{dembx\_OpenMP} & Preconditioned conjugate gradient relaxation \\
\texttt{stress\_OpenMP} & Global/voxel stress-strain calculation \\
\texttt{ppixel\_hdf5}, \texttt{assig} & Microstructure assignment, phase fraction statistics \\
\texttt{rotatestiffnessby} & Crystal orientation to stiffness tensor conversion \\
\texttt{stress\_fullfield\_OpenMP} & Full-field von Mises stress output (HDF5) \\
\bottomrule
\end{tabular}
\end{table}

\begin{figure}[htbp]
  \centering
  \includegraphics[width=0.85\textwidth]{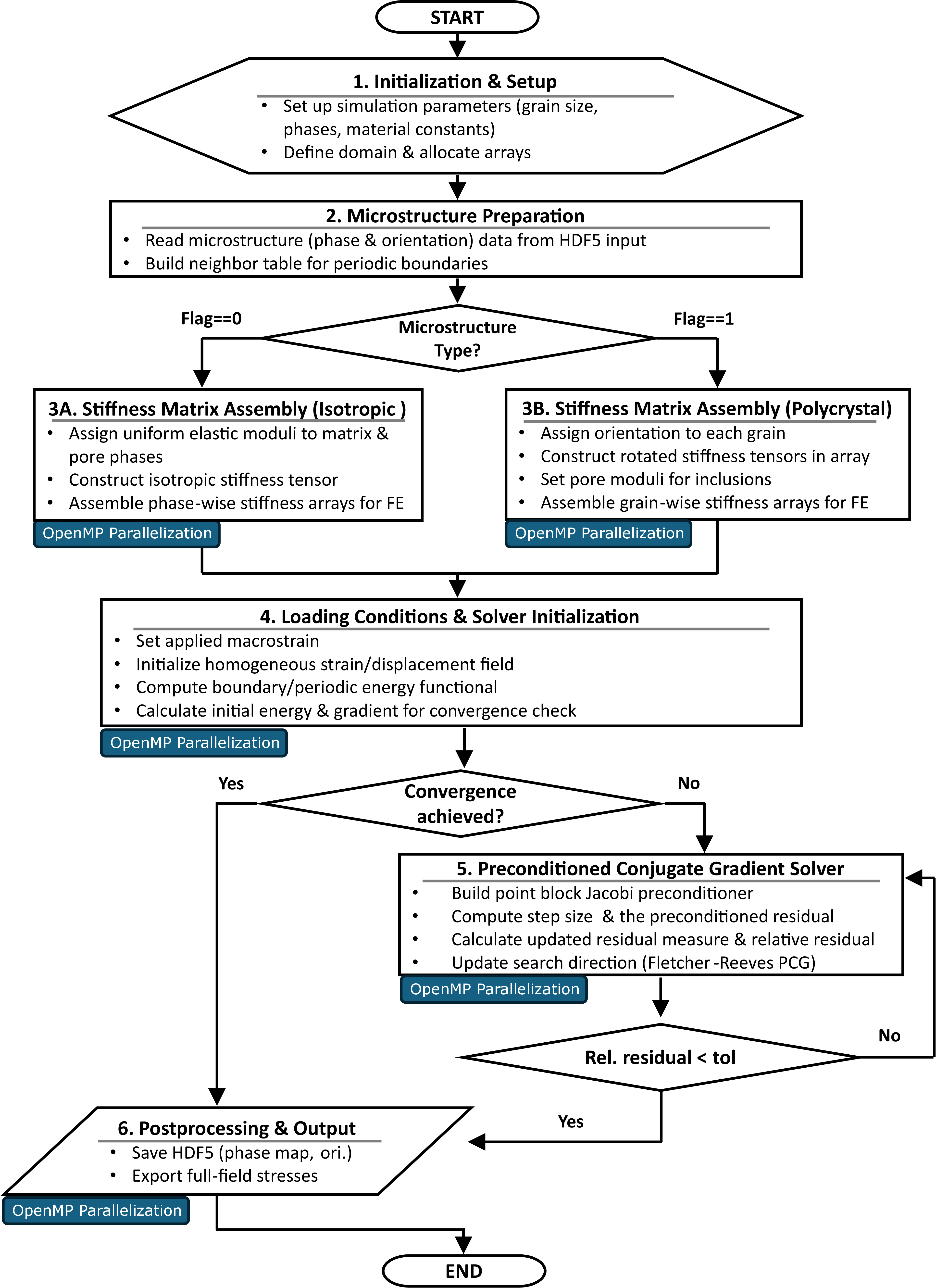}
  \caption{Computational workflow for \texttt{ELAS3D-Xtal}. The algorithm proceeds as follows:
  \textbf{(1--2) Initialization \& Prep:} Simulation parameters are defined and microstructure data is read from HDF5 inputs to build periodic neighbor tables.
  \textbf{(3) Stiffness Assembly:} A decision tree directs OpenMP-parallelized assembly for either isotropic media (\texttt{flag=0}, uniform moduli) or polycrystals (\texttt{flag=1}, rotated stiffness tensors).
  \textbf{(4--5) Solver Loop:} After applying macroscopic strain and initializing energy functionals, the OpenMP-accelerated PCG solver employs point-block Jacobi preconditioning and Fletcher-Reeves updates until convergence.
  \textbf{(6) Output:} Final phase maps and full-field stresses are exported.}
  \label{fig:1}
\end{figure}

The computational workflow, illustrated in Figure~\ref{fig:1}, proceeds through five sequential stages. First, the simulation environment is established by defining global simulation parameters. Second, the initialization phase reads microstructure and crystallographic orientation data from HDF5 files and constructs internal periodic neighbor look-up tables. Third, the assembly phase computes the local stiffness contributions; a decision tree selects between isotropic phases using Lamé constants or polycrystalline phases using the tensor rotation routines described in Section~\ref{sec2:1:1}. Fourth, the loading and solver initialization phase applies the macroscopic strain tensor $\boldsymbol{\epsilon}^{\text{app}}$ to generate the initial affine displacement field $\mathbf{u}^0$ and evaluates the initial system energy and gradient to establish the baseline for convergence monitoring. Lastly, the solution phase executes the preconditioned conjugate gradient algorithm. This stage dominates the runtime, iterating until the force residual satisfies the convergence criterion $||\mathbf{g}||/||\mathbf{g}_0|| < \mathrm{tol}$. Following convergence, the post-processing phase recovers the full-field stresses.

The updated solver, \texttt{ELAS3D-Xtal}, is significantly faster than the original NIST \texttt{ELAS3D} code (Table~\ref{tab:solver_performance}). Using OpenMP, it runs efficiently on multi-core CPUs, reducing runtime for large RVEs. A point-block Jacobi preconditioner improves convergence for materials with strong stiffness contrast, such as porous ceramics and polycrystalline metals, where simple scalar Jacobi scaling performs poorly. The solver is also matrix-free and uses cache-blocked stencil operations, reducing memory usage and enabling larger, more detailed microstructure simulations. Full-field results are estimated in parallel and stored in HDF5 format.

\begin{table}[ht]
\centering
\caption{Comparison of original and upgraded ELAS3D solver capabilities}
\label{tab:solver_performance}
\begin{tabular}{llc}
\toprule
\textbf{Feature} & \textbf{NIST ELAS3D} & \textbf{ELAS3D-Xtal} \\
\midrule
Parallelization & Serial & OpenMP multi-core \\
Preconditioner & None & Block Jacobi \\
Material models & Isotropic & Anisotropic, pore, orientation \\
I/O & Text & Text \& HDF5 \\
Output fields & Limited & Full-field, diagnostics, von Mises \\
\bottomrule
\end{tabular}
\end{table}

\subsection{Mixed polycrystal generation}
\label{sec2:2}
\subsubsection{Grain seeding and spatial acceleration}
\label{sec2:2:1}
The microstructure generation process begins with the initialization of a discrete set of grain nuclei, as detailed in Algorithm~\ref{alg:grain_seeding}. To ensure the synthetic volume matches the physical statistics of laser powder bed fusion (LPBF) stainless steel 316L, the total number of grains, $N_g$, is estimated by dividing the total domain volume by a mean grain volume. This volume is approximated cylindrically using the mean diameter and aspect ratio to ensure a consistent fill density. For each grain $i$, a centroid $\mathbf{s}_i$ is placed at a uniformly random coordinate $(x_i, y_i, z_i)$ within the domain boundaries. 

The grain morphology is defined as a prolate spheroid characterized by a transverse radius $r_{t,i}$ (in the XY plane) and a longitudinal radius $r_{l,i}$ (along the build direction Z). The grain diameter $d_i$ and aspect ratio $\alpha_i$ are sampled from lognormal distributions calibrated to EBSD measurements taken from the three outer orthogonal surfaces~\cite{andani2024mapping,douglas2024influence}:
\begin{equation}
d_i \sim \text{Lognormal}(\mu_d, \sigma_d), \quad
\alpha_i \sim \text{Lognormal}(\mu_\alpha, \sigma_\alpha).
\label{eqn:grain_stats}
\end{equation}
The semi-axes are subsequently computed as $r_{t,i} = d_i/2$ and $r_{l,i} = \alpha_i \cdot r_{t,i}$. This statistical approach to generating synthetic microstructure is consistent with established frameworks for representative volume elements~\cite{quey2011large,groeber2014dream}.

\begin{algorithm}
\caption{Grain Seeding and Spatial Acceleration Structure}
\label{alg:grain_seeding}
\small
\SetAlgoLined
\DontPrintSemicolon

\KwIn{Domain bounds $\mathbf{x}_{\text{vec}}, \mathbf{y}_{\text{vec}}, \mathbf{z}_{\text{vec}}$; Volume $V_{\text{domain}}$\\
Grain statistics: $\mu_d, \sigma_d$ (diameter), $\mu_\alpha, \sigma_\alpha$ (aspect ratio)\\
Grid resolution $G$ for spatial partitioning}

\KwOut{Grain parameters: seeds $\{\mathbf{s}_i\}$, semi-axes $\{r_{t,i}, r_{l,i}\}$\\
Spatial acceleration grid $\mathcal{G}_{ijk}$ (Cell array)}

\BlankLine
\tcp{1: Grain Seeding}
Estimate mean grain volume: $V_{\text{grain}}^{\text{mean}} \leftarrow \pi (\mu_d/2)^2 \cdot (\mu_\alpha \cdot \mu_d)$\;
Estimate grain count: $N_g \leftarrow \lceil V_{\text{domain}} / V_{\text{grain}}^{\text{mean}} \rceil$\;

\BlankLine

\For{$i = 1$ \KwTo $N_g$}{
    \tcp{Sample statistics}
    \Indp
    $d_i \sim \text{Lognormal}(\mu_d, \sigma_d)$; \quad $\alpha_i \sim \text{Lognormal}(\mu_\alpha, \sigma_\alpha)$\;
    \Indm

    \tcp{Compute semi-axes (Transverse \& Longitudinal)}
    \Indp
    $r_{t,i} \leftarrow d_i / 2$; \quad $r_{l,i} \leftarrow \alpha_i \cdot r_{t,i}$\;
    \Indm
    
    \tcp{Generate random seed position}
    \Indp
    $\mathbf{s}_i \leftarrow (x_i, y_i, z_i)$ where $x_i \sim \mathcal{U}[x_{\min}, x_{\max}]$, $y_i \sim \mathcal{U}[y_{\min}, y_{\max}]$, $z_i \sim \mathcal{U}[z_{\min}, z_{\max}]$\;
    \Indm
}

\BlankLine
\tcp{2: Spatial Acceleration Structure}
Compute max global influence: $R_{\max} \leftarrow 1.5 \cdot \max_{i=1,\ldots,N_g}(\max(r_{t,i}, r_{l,i}))$\;

\BlankLine
Define grid edges: $x_{\text{edges}}, y_{\text{edges}}, z_{\text{edges}}$ using $G$ partitions\;

\BlankLine
\ForPar{$idx = 1$ \KwTo $G^3$}{
    Map linear $idx$ to subscript $(i,j,k)$\;
    Get cell bounds: $[x_{\min}^c, x_{\max}^c], [y_{\min}^c, y_{\max}^c], [z_{\min}^c, z_{\max}^c]$\;
    
    \tcp{Identify relevant grains (intersect cell + margin)}
    $\mathcal{G}_{ijk} \leftarrow \left\{ \ell \in \{1,\dots,N_g\} \mid \right.$ \;
    \Indp
    $x_{\min}^c - R_{\max} \leq s_{\ell,x} \leq x_{\max}^c + R_{\max} \ \land$\;
    $y_{\min}^c - R_{\max} \leq s_{\ell,y} \leq y_{\max}^c + R_{\max} \ \land$\;
    $z_{\min}^c - R_{\max} \leq s_{\ell,z} \leq z_{\max}^c + R_{\max})$\;
    \Indm
}

\BlankLine
\Return{$\{\mathbf{s}_i, r_{t,i}, r_{l,i}\}_{i=1}^{N_g}$, $\mathcal{G}_{ijk}$}\;
\end{algorithm}

To mitigate the high computational cost of assigning voxels in large domains, we construct a spatial acceleration structure prior to voxelization. A naive assignment strategy would require the checking of every voxel against every grain on the global list ($O(N_v \cdot N_g)$). To avoid this bottleneck, the domain is partitioned into a coarse, uniform $15 \times 15 \times 15$ cell grid, which can be adjusted via user settings. The algorithm effectively pre-sorts grains into these cells, creating a localized lookup table.

A grain whose center lies in one grid cell can still extend into neighboring cells due to its finite volume. To capture these cross-boundary effects, we define a maximum influence radius, $R_{\max} = 1.5 \cdot \max(r_{\text{all}})$, based on the largest grain dimension in the population. The domain is then scanned cell-by-cell in parallel. The solver builds a local list of candidate grains, $\mathcal{G}_{ijk}$, for each cell. A grain $\ell$ is included if its seed coordinate $\mathbf{s}_{\ell}$ lies inside the cell boundaries expanded outward by $R_{\max}$. This overlap check ensures that, in the subsequent voxel assignment stage (Algorithm~\ref{alg:grain_assignment}), each voxel queries only the grains in its cell’s local candidate list instead of the entire set of grains.

\subsubsection{Parallel grain assignment}
\label{sec2:2:2}
Once the spatial indexing of the grain seeds is completed, the voxel mesh is partitioned into distinct computational blocks to facilitate efficient parallel processing (Algorithm~\ref{alg:grain_assignment}). Within each block, acting as an independent sub-domain, the solver performs spatial filtering to reduce the search space~\cite{ericson2004real}: it queries the previously generated spatial grid index to retrieve only the local subset of relevant grains, $\mathcal{G}_b$, that physically intersect or neighbor the block's bounding box. 

For every voxel $\mathbf{v}$ in position $(x,y,z)$, the solver computes the weighted ellipsoidal distance to the local candidate grain $i \in \mathcal{G}_b$. Using the transverse ($r_{t,i}$) and longitudinal ($r_{l,i}$) semi-axes defined in the seeding stage, the distance metric is given by:
\begin{equation}
D_i^2 = \frac{(x - s_{i,x})^2 + (y - s_{i,y})^2}{r_{t,i}^2} + \frac{(z - s_{i,z})^2}{r_{l,i}^2}.
\label{eqn:ellipsoid_dist}
\end{equation}
The voxel is assigned to the nearest grain that minimizes the squared distance, $\phi(\mathbf{v}) = \arg\min_{i \in \mathcal{G}_b} D_i^2$.

The computational efficiency of this framework is best understood by comparing it to a standard brute-force approach. In a brute-force method, every voxel must calculate the distances to all $N_g$ grains in the domain to find the nearest one. This results in a total operation count proportional to $O_{brute}(N_{\text{voxels}} \cdot N_g)$. For a typical microstructure with $N_g \approx 10,000$ grains, this calculation becomes prohibitively slow.

In contrast, the proposed spatial filtering approach limits search-distance queries to the local subset $\bar{n}_g$ (typically $50\text{--}100$ grains) within the grid cell. The operation count scales as $O_{filtered}(N_{\text{voxels}} \cdot \bar{n}_g)$. The algorithmic speedup is determined by the ratio of total grains to local grains ($N_g / \bar{n}_g$). With $N_g \sim 10^4$ and $\bar{n}_g \sim 10^2$, this yields an intrinsic two orders of magnitude ($100\times$) reduction in computational cost, independent of hardware. When coupled with parallel block processing, this optimization allows for the generation of high-resolution microstructures ($>10^8$ voxels) in minutes rather than hours (see Table~\ref{tab:alg_comparison1}).

\begin{algorithm}
\caption{Parallel Block-Based Grain Assignment}
\label{alg:grain_assignment}
\small
\SetAlgoLined
\DontPrintSemicolon

\KwIn{Voxel mesh dimensions $(N_x, N_y, N_z)$, coordinate vectors $(\mathbf{x}_{\text{vec}}, \mathbf{y}_{\text{vec}}, \mathbf{z}_{\text{vec}})$\\
Grain data: seeds $\{\mathbf{s}_i\}$, semi-axes $\{r_{xy,i}, r_{z,i}\}$ for $i=1,\ldots,N_g$\\
Spatial grid $\{\mathcal{G}_{ijk}\}$ from Algorithm~\ref{alg:grain_seeding}, Block size $B$}

\KwOut{Phase map $\phi \in \mathbb{N}^{N_x \times N_y \times N_z}$ with grain IDs}

\BlankLine
\tcp{3: Parallel Block Processing}
Initialize phase map: $\phi \leftarrow \mathbf{0}_{N_x \times N_y \times N_z}$\;

\BlankLine
Compute block partitioning:\;
\Indp
$n_x \leftarrow \lceil N_x / B \rceil$, $n_y \leftarrow \lceil N_y / B \rceil$, $n_z \leftarrow \lceil N_z / B \rceil$\;
$n_{\text{blocks}} \leftarrow n_x \times n_y \times n_z$\;
\Indm

\BlankLine
\ForPar{$b = 1$ \KwTo $n_{\text{blocks}}$}{
\tcp{A. Convert linear block index to 3D coordinates}
$b_z \leftarrow \lceil b / (n_x \cdot n_y) \rceil$\;
$b_y \leftarrow \lceil (b - (b_z - 1) \cdot n_x \cdot n_y) / n_x \rceil$\;
$b_x \leftarrow b - (b_z - 1) \cdot n_x \cdot n_y - (b_y - 1) \cdot n_x$\;

\BlankLine
\tcp{B. Determine voxel index ranges for this block}
$i_{\min} \leftarrow (b_x - 1) \cdot B + 1$; \quad $i_{\max} \leftarrow \min(b_x \cdot B, N_x)$\;
$j_{\min} \leftarrow (b_y - 1) \cdot B + 1$; \quad $j_{\max} \leftarrow \min(b_y \cdot B, N_y)$\;
$k_{\min} \leftarrow (b_z - 1) \cdot B + 1$; \quad $k_{\max} \leftarrow \min(b_z \cdot B, N_z)$\;

\BlankLine
\tcp{C. Extract block coordinates}
$\mathbf{X}_b, \mathbf{Y}_b, \mathbf{Z}_b \leftarrow \text{meshgrid}(\mathbf{x}_{\text{vec}}[i_{\min}:i_{\max}], \mathbf{y}_{\text{vec}}[j_{\min}:j_{\max}], \mathbf{z}_{\text{vec}}[k_{\min}:k_{\max}])$\;

\BlankLine
\tcp{D. Query spatial grid for relevant grains}
Compute block bounds: $x_{\min}^b, x_{\max}^b, y_{\min}^b, y_{\max}^b, z_{\min}^b, z_{\max}^b$\;

Map block to grid cells: $g_x^{\min}, g_x^{\max}, g_y^{\min}, g_y^{\max}, g_z^{\min}, g_z^{\max}$\;

$\mathcal{G}_b \leftarrow \bigcup_{i=g_x^{\min}}^{g_x^{\max}} \bigcup_{j=g_y^{\min}}^{g_y^{\max}} \bigcup_{k=g_z^{\min}}^{g_z^{\max}} \mathcal{G}_{ijk}$

\BlankLine
\tcp{E. Initialize block results}
$\phi_b \leftarrow \mathbf{0}_{\text{block size}}$; \quad $D_{\min}^2 \leftarrow \infty \cdot \mathbf{1}_{\text{block size}}$

\BlankLine
\tcp{F. Assign each voxel to the nearest grain}
\For{each grain $\ell \in \mathcal{G}_b$}{
Compute displacement fields:\;
\Indp
$\Delta \mathbf{X} \leftarrow \mathbf{X}_b - s_{\ell,x}$, $\Delta \mathbf{Y} \leftarrow \mathbf{Y}_b - s_{\ell,y}$, $\Delta \mathbf{Z} \leftarrow \mathbf{Z}_b - s_{\ell,z}$
\Indm

\BlankLine
Compute weighted ellipsoidal distance: $D_{\ell}^2 \leftarrow \frac{(\Delta \mathbf{X})^2 + (\Delta \mathbf{Y})^2}{r_{t,\ell}^2} + \frac{(\Delta \mathbf{Z})^2}{r_{l,\ell}^2}$\;

\BlankLine
Update nearest phase map: $\text{mask} \leftarrow (D_{\ell}^2 < D_{\min}^2)$\;
        $\phi_b[\text{mask}] \leftarrow \ell$; \quad $D_{\min}^2[\text{mask}] \leftarrow D_{\ell}^2[\text{mask}]$\;

}

\BlankLine
\tcp{G. Store block result}
$\phi[i_{\min}:i_{\max}, j_{\min}:j_{\max}, k_{\min}:k_{\max}] \leftarrow \phi_b$\;

}
\BlankLine
\Return{$\phi$}\;
\end{algorithm}

\begin{table}[ht]
\centering
\caption{Computational characteristics of the two-stage microstructure generation algorithm. The spatial filtering approach yields a significant reduction in comparisons per voxel compared to a brute-force baseline.}
\label{tab:alg_comparison1}
\small
\begin{tabular}{@{}lcc@{}}
\toprule
\textbf{Characteristic} & \textbf{Algorithm~\ref{alg:grain_seeding}} & \textbf{Algorithm~\ref{alg:grain_assignment}} \\
\midrule
Primary function & Grain seeding \& indexing & Voxel-to-grain assignment \\
Parallelization strategy & Grid cells ($15^3$) & Voxel blocks ($\sim$10$^4$) \\
Search complexity & $O(N_g \cdot G^3)$ & $O(N_v \cdot \bar{n}_g)$ \\
Avg. comparisons / voxel & N/A & $\sim$50 (vs. $\sim$10,000 Brute Force) \\
Typical runtime & $<$1 second & $\sim$5 minutes \\
Speedup factor & N/A & $>100\times$ (via Spatial Filtering) \\
\bottomrule
\end{tabular}
\end{table}

\subsubsection{Pore integration and orientation assignment}
\label{sec2:2:3}
After the polycrystal tessellation, defect populations derived from X-ray computed tomography (XCT) are superimposed onto the voxel grid. This is implemented as a boolean overwrite operation detailed in Algorithm~\ref{alg:pore_integration}. For every pore $p$ with center $\mathbf{c}_p$ and radius $r_p$, the squared Euclidean distance $D^2(\mathbf{x}) = \|\mathbf{x} - \mathbf{c}_p\|^2$ is computed for local voxels relative to the pore center. Any voxel satisfying the spherical radius condition $D^2(\mathbf{x}) \leq r_p^2$ is reassigned a distinct pore phase ID ($\phi(\mathbf{x}) = N_g + 1$), carving voids out of the existing grain structure.

Crystallographic orientations are assigned to each grain $k \in {1, \dots, N_g}$ to define local elastic anisotropy. To ensure compatibility with the finite element solver, these orientations are parameterized as Rodrigues vectors, $\boldsymbol{\rho}_k = \tan(\theta_k/2) \hat{\mathbf{n}}_k$. Two assignment modes shown in Algorithm~\ref{alg:orientation_assignment} are implemented to replicate typical additively manufactured conditions: (1) a fiber texture for strong epitaxial grain growth, and (2) a random texture characteristic of the weak crystallographic alignment observed in Lack-of-Fusion (LoF) SS316L.

To replicate the strong solidification textures characteristic of LPBF processes, a fiber texture assignment mode is implemented. In this scheme, grain orientations are constructed by first aligning the crystal axis $\langle 101 \rangle$ with the sample build direction ($X_3$). A random rotational spin $\phi \in [0, 2\pi]$ is applied around the $X_3$-axis, followed by a Gaussian wobble rotation sampled from a normal distribution $\mathcal{N}(0, \sigma_{\text{spread}}^2)$ to mimic the experimental texture dispersion observed in the as-built specimens.

For random texture validation studies, assuming statistical isotropy, the framework includes a random orientation mode. In this approach, orientations are sampled from the uniform Haar measure on the rotation group $SO(3)$~\cite{morawiec2003orientations}. This is achieved numerically using Shoemake's subgroup algorithm~\cite{shoemake1992uniform}, where unit quaternions $\mathbf{q}$ are generated from three independent uniform random variables $u_1, u_2, u_3 \sim \mathcal{U}[0,1]$:
\begin{equation}
    \mathbf{q} = \left[ \sqrt{1-u_1}\sin(2\pi u_2), \quad \sqrt{1-u_1}\cos(2\pi u_2), \quad \sqrt{u_1}\sin(2\pi u_3), \quad \sqrt{u_1}\cos(2\pi u_3) \right]^T
\end{equation}
The resulting quaternions $\mathbf{q}$ are converted to Rodrigues vectors $\boldsymbol{\rho}$ for solver input, and inverse pole figure (IPF) coloring maps are generated via the MTEX toolbox~\cite{bachmann2010texture} to visualize the spatial distribution of crystallographic textures. 

\begin{algorithm}[t]
\caption{Pore Integration via Spherical Overlap}
\label{alg:pore_integration}
\small
\SetAlgoLined
\DontPrintSemicolon

\KwIn{Phase map $\phi$ from Algorithm~\ref{alg:grain_assignment}\\
Voxel mesh coordinates $(X, Y, Z) \in \mathbb{R}^{N_x \times N_y \times N_z}$\\
XCT pore data: centers $\{\mathbf{c}_p = (x_p, y_p, z_p)\}_{p=1}^{N_p}$, radii $\{r_p\}_{p=1}^{N_p}$\\
Domain bounds: $[x_{\min}, x_{\max}] \times [y_{\min}, y_{\max}] \times [z_{\min}, z_{\max}]$\\
Number of grains: $N_g$}

\KwOut{Updated phase map $\phi$ with pore phase assignments}

\BlankLine
\tcp{4: Pore Region Assignment}
Initialize pore counter: $n_{\text{active}} \leftarrow 0$\;
Assign pore phase ID: $\text{ID}_{\text{pore}} \leftarrow N_g + 1$\;

\BlankLine
\tcp{Filter and map pores within domain}
\For{$p = 1$ \KwTo $N_p$}{
\tcp{Compute squared distance from pore center $(x_p, y_p, z_p)$}
$D^2 \leftarrow (X - x_p)^2 + (Y - y_p)^2 + (Z - z_p)^2$\;

\BlankLine
Create mask for voxels inside sphere:\;
$\mathcal{M}_p \leftarrow \{(i,j,k) \mid D^2(i,j,k) \leq r_p^2\}$\;

\BlankLine
\tcp{Overwrite grain assignments with pore phase}
\If{$|\mathcal{M}_p| > 0$}{
$\phi[\mathcal{M}_p] \leftarrow \text{ID}_{\text{pore}}$; $n_{\text{active}} \leftarrow n_{\text{active}} + 1$\;
}
}

\BlankLine
\tcp{Update phase count}
$n_{\text{phases}} \leftarrow N_g + 1$

\BlankLine
\Return{$\phi$, $n_{\text{phases}}$}\;
\end{algorithm}

\begin{algorithm}
\caption{Crystallographic Orientation Assignment (Texture \& Random)}
\label{alg:orientation_assignment}
\small
\SetAlgoLined
\DontPrintSemicolon

\KwIn{Number of grains $N_g$\\
Texture type: $\tau \in \{\text{`textured'}, \text{`random'}\}$\\
\textit{If textured:} Crystal axis $\mathbf{c} = [1, 0, 1]^T$, Sample axis $\mathbf{s} = [0, 0, 1]^T$ (Build Dir.),\\
\hspace{1.9cm} Texture spread $\sigma_{\text{spread}}$ (in degrees)}

\KwOut{Orientation vector $\boldsymbol{\rho} \in \mathbb{R}^{3N_g}$ (Rodrigues parametrization)}

\BlankLine
\tcp{5: Crystallographic Orientation Assignment}
Initialize: $\boldsymbol{\rho} \leftarrow \mathbf{0}_{3N_g}$\;

\BlankLine
\uIf{$\tau = \text{`textured'}$}{
\tcp{Fiber Texture Generation}
\tcp{A: Base Rotation, Align Crystal [101] to Sample $X_3$}
Normalize: $\hat{\mathbf{c}} \leftarrow \mathbf{c} / \|\mathbf{c}\|$, $\hat{\mathbf{s}} \leftarrow \mathbf{s} / \|\mathbf{s}\|$\;
Compute rotation axis: $\mathbf{v} \leftarrow \hat{\mathbf{c}} \times \hat{\mathbf{s}}$; \quad $s \leftarrow \|\mathbf{v}\|$; \quad $c \leftarrow \hat{\mathbf{c}} \cdot \hat{\mathbf{s}}$\;

\lIf{$s < 10^{-8}$}{$\mathbf{R}_{\text{base}} \leftarrow \mathbf{I}_3$}
\lElse{$\mathbf{R}_{\text{base}} \leftarrow \mathbf{I}_3 + [\mathbf{v}]_{\times} + [\mathbf{v}]_{\times}^2 \frac{1-c}{s^2}$ \tcp*[h]{$[\mathbf{v}]_{\times} = \text{skew}(\mathbf{v})$}}

\BlankLine
\For{$k = 1$ \KwTo $N_g$}{
\tcp{B: Random Spin (Fiber Ring) \& Gaussian Wobble (Spread)}
$\phi \sim \mathcal{U}[0, 2\pi]$; \quad $\mathbf{R}_{\text{spin}} \leftarrow \begin{bmatrix} \cos\phi & -\sin\phi & 0 \\ \sin\phi & \cos\phi & 0 \\ 0 & 0 & 1 \end{bmatrix}$\;

$\alpha \sim \mathcal{U}[0, 2\pi]; \quad \theta_{\text{wob}} \sim \mathcal{N}(0, \sigma_{\text{spread}}^2)$\;
$\mathbf{R}_{\text{wobble}} \leftarrow \text{RodriguesRot}([\cos\alpha, \sin\alpha, 0]^T, \theta_{\text{wob}})$\;

\BlankLine
\tcp{C: Combine Rotations \& Convert to Rodrigues vector}
$\mathbf{R}_{\text{total}} \leftarrow \mathbf{R}_{\text{spin}} \cdot \mathbf{R}_{\text{wobble}} \cdot \mathbf{R}_{\text{base}}$\;

$\boldsymbol{\rho}[3(k-1)+1 : 3k] \leftarrow \text{MatrixToRodrigues}(\mathbf{R}_{\text{total}})$\;
}
}
\ElseIf{$\tau = \text{`random'}$}{
    \tcp{Uniform Random Orientation (Haar Measure)}
    \For{$k = 1$ \KwTo $N_g$}{
        \tcp{A: Generate Uniform Quaternion (Shoemake's Method)}
        $u_1, u_2, u_3 \sim \mathcal{U}[0, 1]$\;
        $q_w, q_x \leftarrow \sqrt{1-u_1} \sin(2\pi u_2), \ \sqrt{1-u_1} \cos(2\pi u_2)$\;
        $q_y, q_z \leftarrow \sqrt{u_1} \sin(2\pi u_3), \ \sqrt{u_1} \cos(2\pi u_3)$\;
        
        \tcp{B: Convert to Rodrigues \& Store}
        \lIf{$|q_w| < 10^{-8}$}{$\boldsymbol{\rho}_k \leftarrow [q_x, q_y, q_z]^T \times 10^9$}
        \lElse{$\boldsymbol{\rho}_k \leftarrow [q_x, q_y, q_z]^T / q_w$}
        
        $\boldsymbol{\rho}[3k-2 : 3k] \leftarrow \boldsymbol{\rho}_k$\;
    }
}

\BlankLine
\Return{$\boldsymbol{\rho}$}\;

\end{algorithm}

\begin{figure}[ht]
  \centering
  \includegraphics[width=1.0\textwidth]{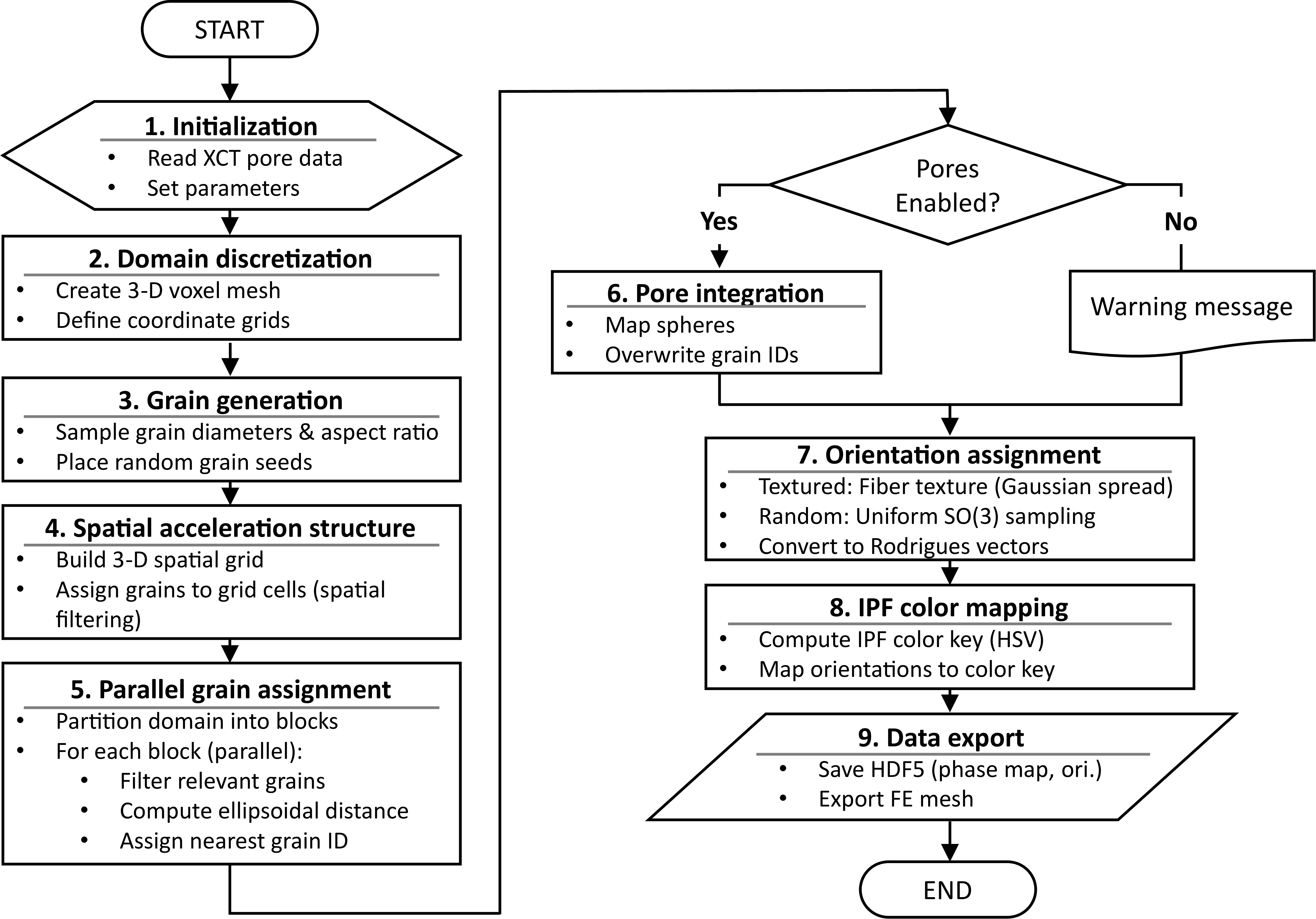}
  \caption{Computational workflow for microstructure generation. The algorithm integrates experimental pore data from XCT scans with statistically representative grain morphologies. Spatial filtering and parallel block processing enable efficient voxel-to-grain assignment. Pore integration overwrites grain assignments with spherical defect regions. Crystallographic orientations are assigned using either fiber texture or random sampling.}
  \label{fig:2}
\end{figure}

\begin{figure}[ht]
  \centering
  \includegraphics[width=1.0\textwidth]{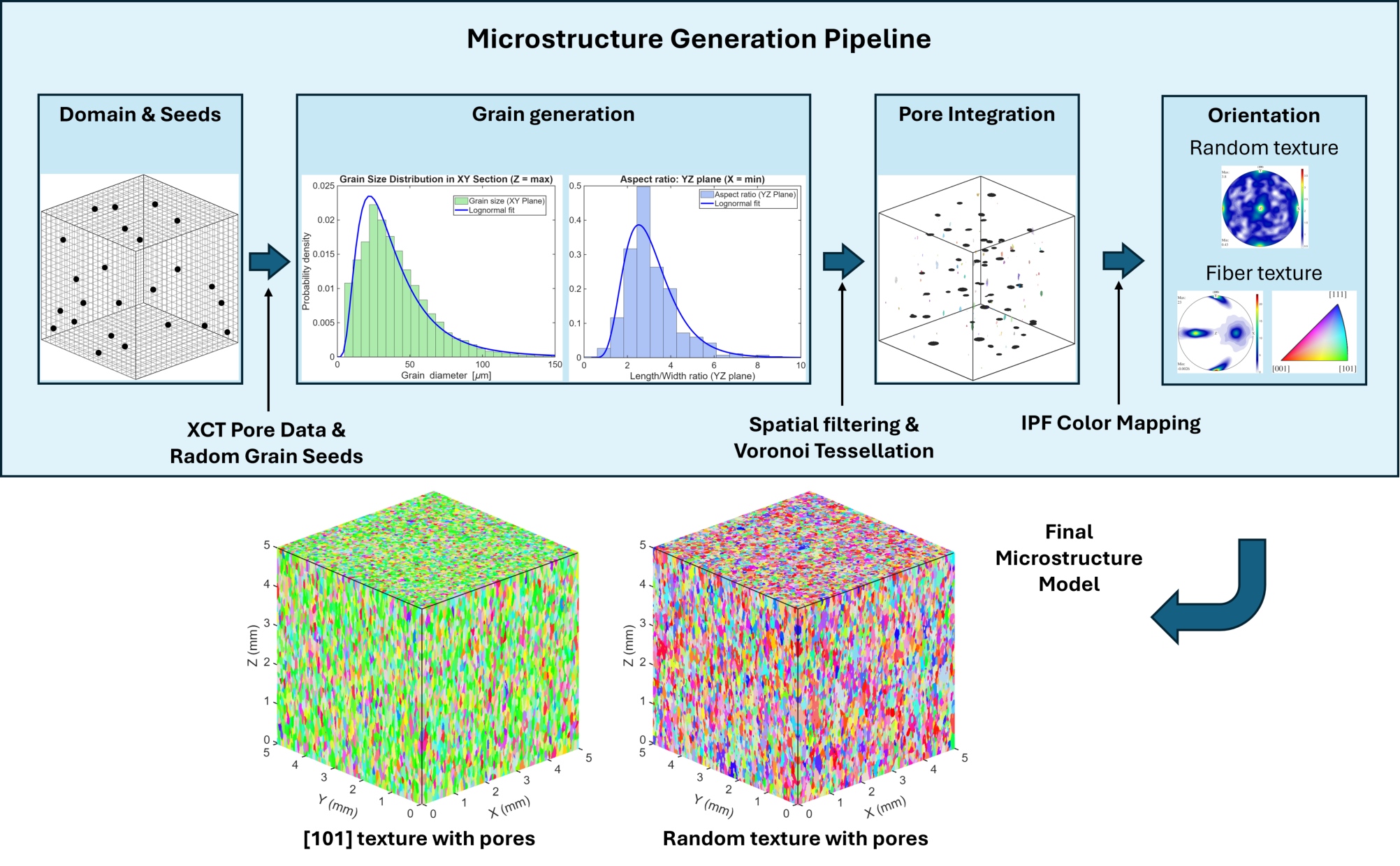}
  \caption{Schematic of the microstructure reconstruction framework. The pipeline features: (1) direct integration of pore data from XCT scans; (2) efficient parallel processing via spatial filtering; and (3) flexible orientation assignment allowing for both uniform SO(3) sampling and Gaussian-spread fiber textures.}
  \label{fig:3}
\end{figure}

The workflow for microstructure generation is described in Figures~\ref{fig:2} and~\ref{fig:3}, and detailed stepwise in Algorithms~\ref{alg:grain_seeding}, \ref{alg:grain_assignment}, \ref{alg:pore_integration}, and~\ref{alg:orientation_assignment}. This computational framework executes in three distinct phases: tessellation, defect integration, and crystallographic assignment. 
The process begins by establishing a voxelized domain populated with grain seeds. As detailed in Algorithms~\ref{alg:grain_seeding} and~\ref{alg:grain_assignment}, the spatial domain is partitioned using a parallel block-based tessellation (Figure~\ref{fig:2}, Step 5). To optimize computational efficiency, the solver employs a spatial filtering technique (Figure~\ref{fig:2}, Step 4) that limits distance calculations to relevant local seeds, thereby significantly reducing overhead for large-scale meshes. 

Once the grain morphology is generated, the defect population is added directly to the voxel grid (Algorithm\ref{alg:pore_integration}). As shown in Figure~\ref{fig:2} (Step 1), pores measured from XCT (centroid locations and radii) are inserted by overwriting the grain phase ID of any voxel inside a pore with distinct pore phase IDs. This Boolean overwrite produces a final model that preserves the polycrystalline structure while explicitly representing the experimentally observed defects in Figure~\ref{fig:2} (Step 6).

Finally, the solver assigns a crystallographic orientation to each grain (Figure\ref{fig:2}, Step 7; Algorithm~\ref{alg:orientation_assignment}). Depending on the simulation requirements, orientations are generated either (i) as a random texture, sampled uniformly from the Haar measure on $SO(3)$, or (ii) as a prescribed fiber texture aligned with the build direction. This final step produces a fully populated, statistically representative volume element that is ready for finite element analysis.

\section{Results}
\label{sec3}
\subsection{Physical benchmark: the Eshelby inclusion problem}
\label{sec3:1}
\begin{figure}[htbp]
  \centering
  \includegraphics[width=1.0\textwidth]{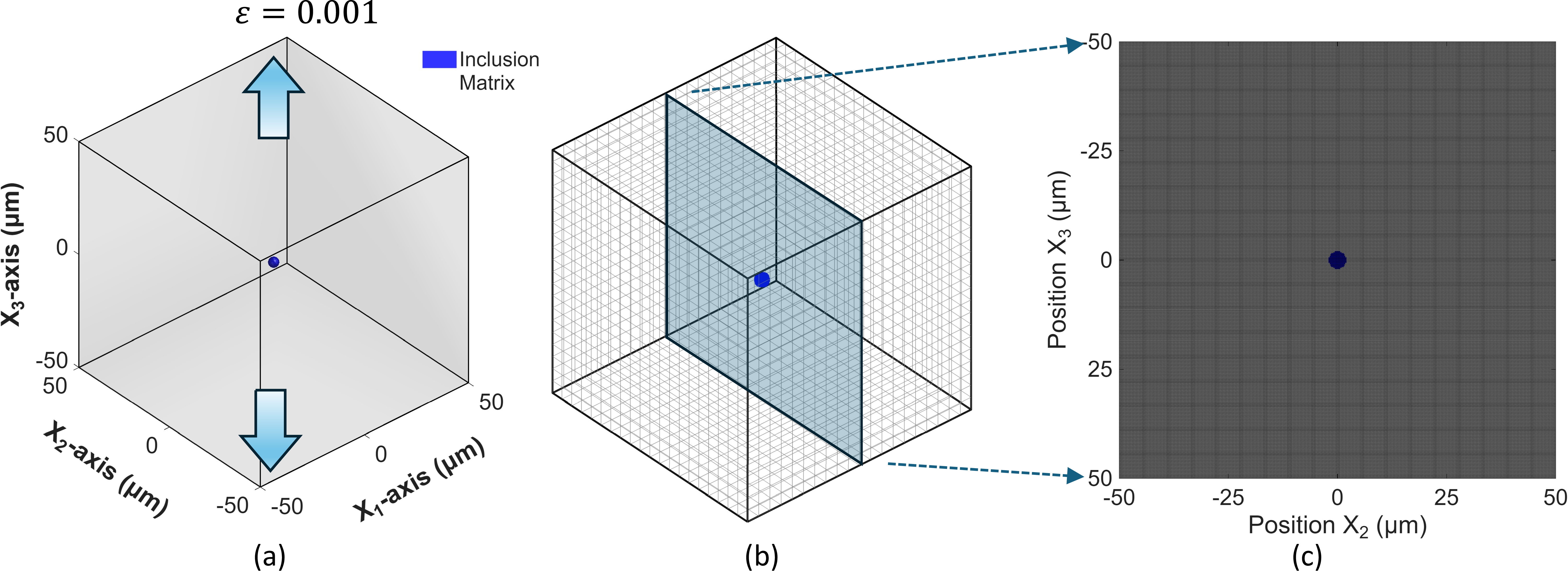}
  \caption{Simulation setup for the 3D Eshelby inclusion problem under macroscopic strain loading along the $X_3$-direction. Material properties are defined for an elastic matrix ($E_m=194$ GPa, $\nu_m = 0.2934$, colored white) surrounding a zero-stiffness void inclusion ($E_{in}=0.0$ GPa, $\nu_{in} = 0.0$, colored blue).
  \textbf{(a)} 3D phase map visualization of the domain.
  \textbf{(b)} Voxelized representation utilizing filled cubes to emphasize the discrete computational grid.
  \textbf{(c)} 2D cross-sectional view taken at the central $X_1$-plane, illustrating the phase distribution overlaid with the finite element mesh grid.}
  \label{fig:4} 
\end{figure}

To validate the mechanical accuracy and evaluate the computational performance of the preconditioned conjugate gradient (PCG) solver, the three-dimensional Eshelby inclusion problem is simulated. The computational setup is illustrated in Figure~\ref{fig:4}. The domain consists of a spherical void (colored blue) located in the center of an isotropic matrix (colored white), as shown in Figure~\ref{fig:4}(a). To approximate the infinite-domain assumption of the analytical Eshelby solution and mitigate the interaction effects from the periodic boundary conditions inherent to the \texttt{ELAS3D-Xtal} solver, the physical domain size is set to $50 \times 50 \times 50$ $\mu\text{m}^3$ with an inclusion radius of $2$ $\mu\text{m}$. The material parameters for the matrix correspond to the standard elastic properties: Young’s modulus $E_m = 194$ GPa and Poisson’s ratio $\nu_m = 0.2934$. The macroscopic domain is subjected to a tensile strain of $0.1\%$ along the $X_3$-direction, which remains within the elastic range. The domain is discretized into a grid of $500 \times 500 \times 500$ voxels, as shown in Figure~\ref{fig:4}(b). A cross-sectional view of the central $x_1$-plane is provided in Figure~\ref{fig:4}(c).

\begin{figure}[htbp]
  \centering
  \includegraphics[width=1.0\textwidth]{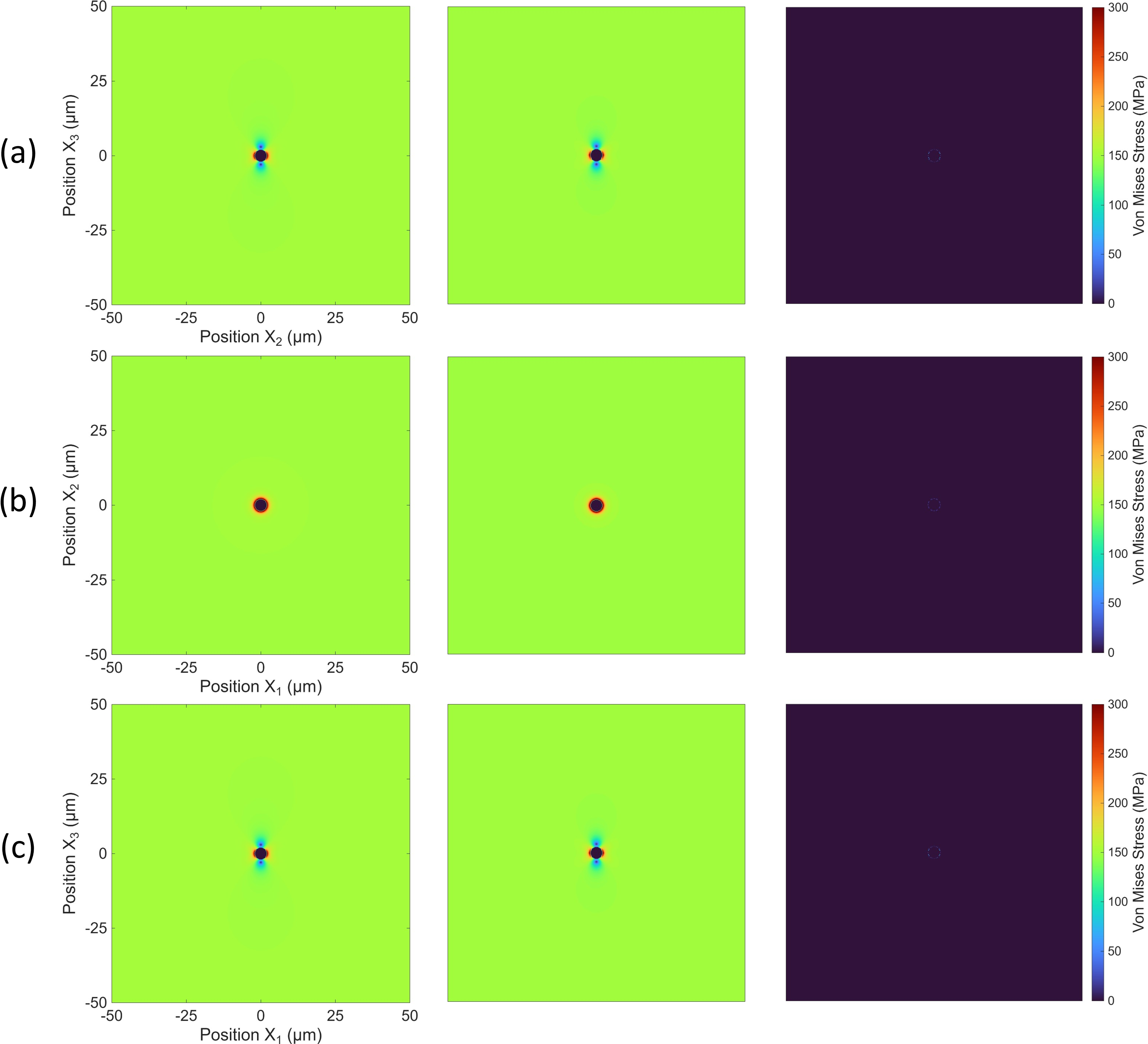}
  \caption{Validation of the numerical solver against the analytical Eshelby solution using a high-resolution $500^3$ mesh. The columns display the analytical solution (left), the numerical PCG solution (middle), and the difference field (right). The rows correspond to central cross-sections in three orthogonal planes:
  \textbf{(a)} the $X_2-X_3$ plane at the center of $X_1$;
  \textbf{(b)} the $X_1-X_2$ plane at the center of $X_3$; and
  \textbf{(c)} the $X_1-X_3$ plane at the center of $X_2$.}
  \label{fig:5}
\end{figure}

The numerical solutions obtained using the PCG solver, as shown in Figure~\ref{fig:5}, are compared with the elastic analytical solution to the Eshelby inclusion problem, which was implemented in MATLAB~\cite{ju1999novel,shao2023linear}. The von Mises stress fields are mapped onto three orthogonal cross-sections through the center of the domain: (a) the $X_2-X_3$ plane at the center of $X_1$, (b) the $X_1-X_2$ plane at the center of $X_3$, and (c) the $X_1-X_3$ plane at the center of $X_2$. In Figure~\ref{fig:5}, the columns show the analytical solution, the numerical solution, and the difference field from left to right, respectively.

Stress concentrations are observed at the solid-void interface, with the stress magnitude rapidly decaying with increasing distance from the void center. Due to symmetry, the von Mises stress mappings in the $X_2-X_3$ and $X_1-X_3$ planes yield identical results. In contrast, the $X_1-X_2$ plane exhibits a high-stress density that forms a ring shape around the void. The difference maps shown in Figure~\ref{fig:5} indicate excellent agreement throughout the domain, with minor deviations confined to the matrix-inclusion interface. These deviations are attributed to the voxelized, zig-zag mesh representation of the spherical boundary. A detailed mesh sensitivity analysis demonstrating the solver's convergence and stability across resolutions from $100^3$ to $500^3$ is provided in~\ref{app1}.

\begin{figure}[htbp]
  \centering
  \includegraphics[width=1.0\textwidth]{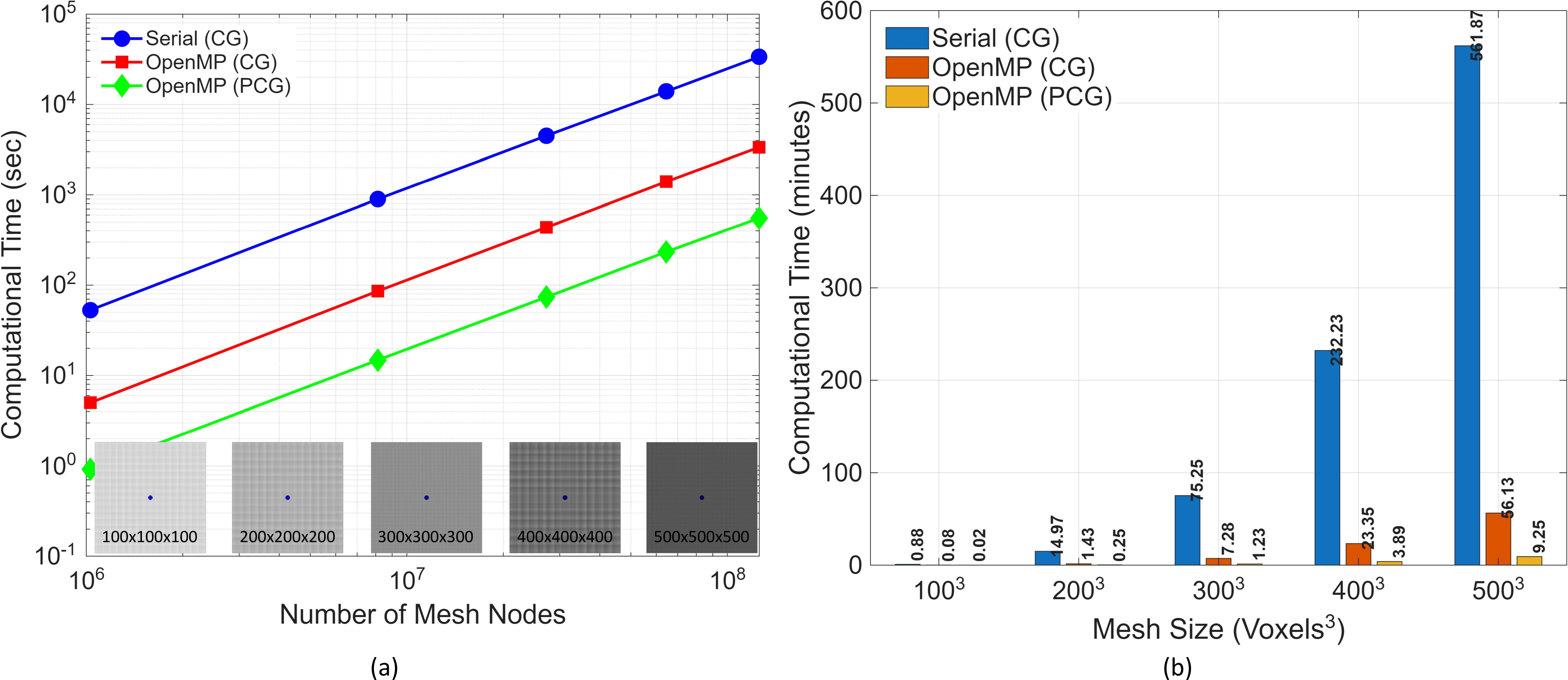}
  \caption{Computational performance scaling for the 3-D FEM simulation framework (\texttt{ELAS3D-Xtal}). (a) The log-log plot illustrates the relationship between mesh density and computational time across five distinct grid resolutions ranging from $100^3$ to $500^3$. Three solver strategies are compared: 
  (Blue circle) the baseline Serial Conjugate Gradient solver; 
  (Red square) the OpenMP-parallelized CG solver; and 
  (Green diamond) the OpenMP-parallelized Preconditioned Conjugate Gradient solver. (b) Grouped bar chart illustrating execution times.}
  \label{fig:6}
\end{figure}

The performance of the proposed solvers is evaluated on voxel-based meshes ranging from $100^3$ to $500^3$ elements, using a convergence tolerance of $10^{-8}$. Table~\ref{tab:performance_summary} reports the computational metrics for the serial conjugate gradient, OpenMP-parallelized CG, and OpenMP-parallelized Preconditioned CG solvers. Figure~\ref{fig:6} illustrates the mesh scaling performance, demonstrating that the OpenMP PCG implementation offers superior efficiency and significantly reduces runtime by orders of magnitude compared to the serial approach for large-scale systems. The log-log plot in Figure~\ref{fig:6}(a) reveals a linear scaling behavior. For the $500^3$ voxel case, the computational time was approximately 561 minutes for the serial CG solver, compared to 56 minutes for the OpenMP CG solver and 9 minutes for the OpenMP PCG solver. The absolute speedup is defined as:
\begin{equation}
S = \frac{T_{\text{serial}}}{T_{\text{parallel}}},
\label{eq:speedup}
\end{equation}
where $T_{\text{serial}}$ denotes the execution time of the serial CG solver, and $T_{\text{parallel}}$ is the execution time of the corresponding parallel implementation on an Intel Core Ultra 9 285K CPU, respectively. The resulting speedups are summarized in Table~\ref{tab:performance_summary}. Across all mesh sizes, OpenMP parallelization yields a consistent speedup of approximately $10\times$, while incorporating preconditioning further increases the speedup to approximately $53$--$61\times$. In addition to absolute speedup, the relative runtime reduction is computed as:
\begin{equation}
R = \frac{T_{\text{serial}} - T_{\text{parallel}}}{T_{\text{serial}}} \times 100\%.
\label{eq:reduction}
\end{equation}
As shown in Table~\ref{tab:performance_summary}, the OpenMP CG solver reduces the runtime by approximately $90\%$, while the OpenMP PCG solver achieves a reduction of about $98\%$ relative to the serial baseline. Notably, these improvements remain nearly mesh-independent, indicating robust solver scalability for large voxel-based systems.

\begin{table}[ht]
\centering
\caption{Comprehensive computational performance summary, including execution time (in seconds), absolute speedup ($S$), and relative runtime reduction ($R$) compared to the Serial CG baseline.}
\label{tab:performance_summary}
\setlength{\tabcolsep}{4pt} 
\begin{tabular}{l c ccc ccc}
\toprule
& \textbf{Serial CG} & \multicolumn{3}{c}{\textbf{OpenMP CG}} & \multicolumn{3}{c}{\textbf{OpenMP PCG}} \\
\cmidrule(lr){2-2} \cmidrule(lr){3-5} \cmidrule(lr){6-8}
Mesh Size & Time (s) & Time (s) & Speedup & Red. (\%) & Time (s) & Speedup & Red. (\%) \\
\midrule
$100^3$ & 53 & 5 & $10.60\times$ & 90.6 & 1 & $53.00\times$ & 98.1 \\
$200^3$ & 898 & 86 & $10.44\times$ & 90.4 & 15 & $59.87\times$ & 98.3 \\
$300^3$ & 4,515 & 437 & $10.33\times$ & 90.3 & 73 & $61.85\times$ & 98.4 \\
$400^3$ & 13,934 & 1,401 & $9.95\times$ & 89.9 & 233 & $59.80\times$ & 98.3 \\
$500^3$ & 33,712 & 3,368 & $10.01\times$ & 90.0 & 555 & $60.74\times$ & 98.4 \\
\bottomrule
\end{tabular}
\end{table}

\subsection{Effect of elastic anisotropy on defect-scale stress fields}
\label{sec3:2}
\begin{figure}[ht]
    \centering
    \includegraphics[width=0.8\textwidth]{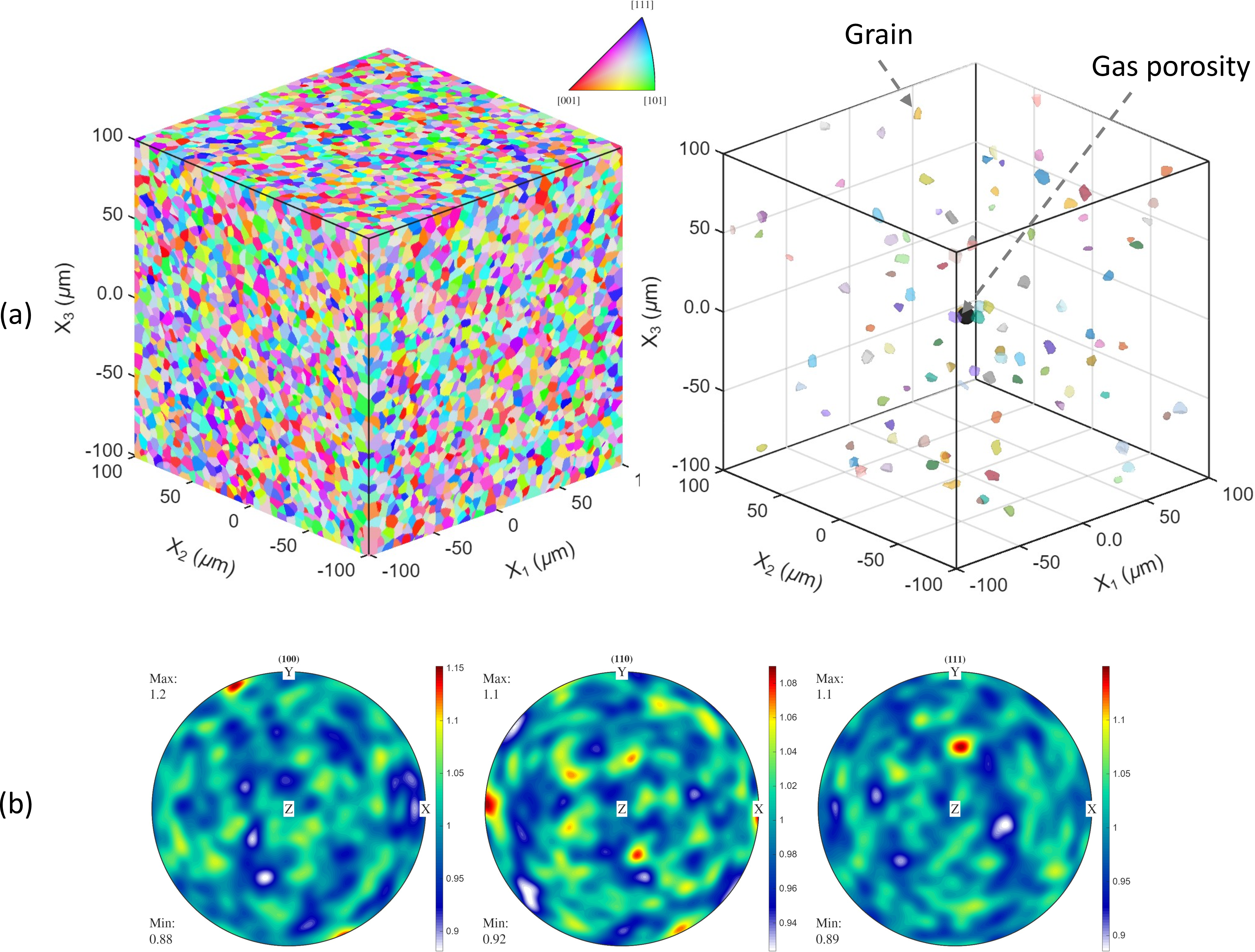}
    \caption{Computational model setup: (a) 3D microstructure ($200^3 \mu m^3$) containing 81,488 equiaxed grains (mean diameter 5 $\mu m$) colored by inverse pole figure orientation, with a central 10 $\mu m$ spherical void. (b) Pole figures demonstrating the random texture distribution for (100), (110), and (111) orientations.}
    \label{fig:7}
\end{figure}

To distinguish the effect of crystalline anisotropy from bulk stiffness on stress concentrations, a sensitivity analysis was performed using the Zener Anisotropy Ratio ($A = 2C_{44}/(C_{11}-C_{12})$). The computational domain, as shown in Figure~\ref{fig:7}, consisted of a $200 \times 200 \times 200$ $\mu m^3$ polycrystal containing 81,488 equiaxed grains generated with a lognormal grain-size distribution and assigned a random crystallographic texture as described in Section~\ref{sec2:2}. 

The grain morphology was characterized by a mean grain size of 5 $\mu m$ with a standard deviation of 0.5 $\mu m$ (10\% of the mean) in the $X_1$-$X_2$ plane. To ensure an equiaxed structure, the mean aspect ratios in the $X_1$-$X_3$ and $X_2$-$X_3$ planes were set to 1.0 with a standard deviation of 0.1. A single spherical gas porosity of radius 5 $\mu m$ was located in the geometric center, as shown in Figure~\ref{fig:7}(a). The domain was discretized into a high-resolution voxel grid of $800^3$ voxels to capture sharp gradients at grain boundaries. Boundary conditions were applied to simulate a macroscopic tensile strain of $\epsilon_{33} = 0.1\%$ ($0.001$) along the $X_3$-direction. The constituent grains were assigned a random crystallographic texture as depicted in Figure~\ref{fig:7}(b).

To facilitate a valid comparison, a constant-stiffness methodology was adopted. For materials with cubic symmetry, the bulk modulus ($K$) and the shear modulus ($G$), calculated using the Voigt average, are defined as follows: \begin{equation} K = \frac{C_{11} + 2C_{12}}{3}, \quad G = \frac{C_{11} - C_{12} + 3C_{44}}{5}. \end{equation}
The baseline material (Case 3, $A=1.0$) was defined with $K=156.64$ GPa and $G=75.08$ GPa. For all other cases, $C_{11}, C_{12},$ and $C_{44}$ were determined by holding $K$ and $G$ fixed at these baseline values and solving the corresponding Zener anisotropy ratio $A$, which ranges from $0.5$ to $5.0$. This approach isolates the influence of texture contrast, ensuring that any observed changes in the stress field are attributable solely to anisotropy rather than to variations in average material stiffness. The resulting data set is detailed in Table \ref{tab:material_params}.

\begin{table}[htbp]
    \centering
    \caption{Material parameters used for the anisotropy sensitivity analysis.}
    \label{tab:material_params}
    \begin{tabular}{c c c c c}
        \hline
        Case ID & Anisotropy Factor ($A$) & $C_{11}$ (GPa) & $C_{12}$ (GPa) & $C_{44}$ (GPa) \\
        \hline
        1 & 0.50 & 300.04 & 84.95 & 53.77 \\
        2 & 0.75 & 274.65 & 97.64 & 66.38 \\
        3 & 1.00 (Baseline) & 256.75 & 106.59 & 75.08 \\      
        4 & 1.50 & 232.07 & 118.93 & 84.85 \\
        5 & 2.00 & 216.52 & 126.70 & 89.82 \\
        6 & 3.00 & 198.66 & 135.64 & 94.53 \\
        7 & 4.00 & 188.74 & 140.60 & 96.28 \\
        8 & 5.00 & 182.38 & 143.78 & 96.50 \\
        \hline
    \end{tabular}
\end{table}

\begin{figure}[htbp]
    \centering
    \includegraphics[width=1.0\textwidth]{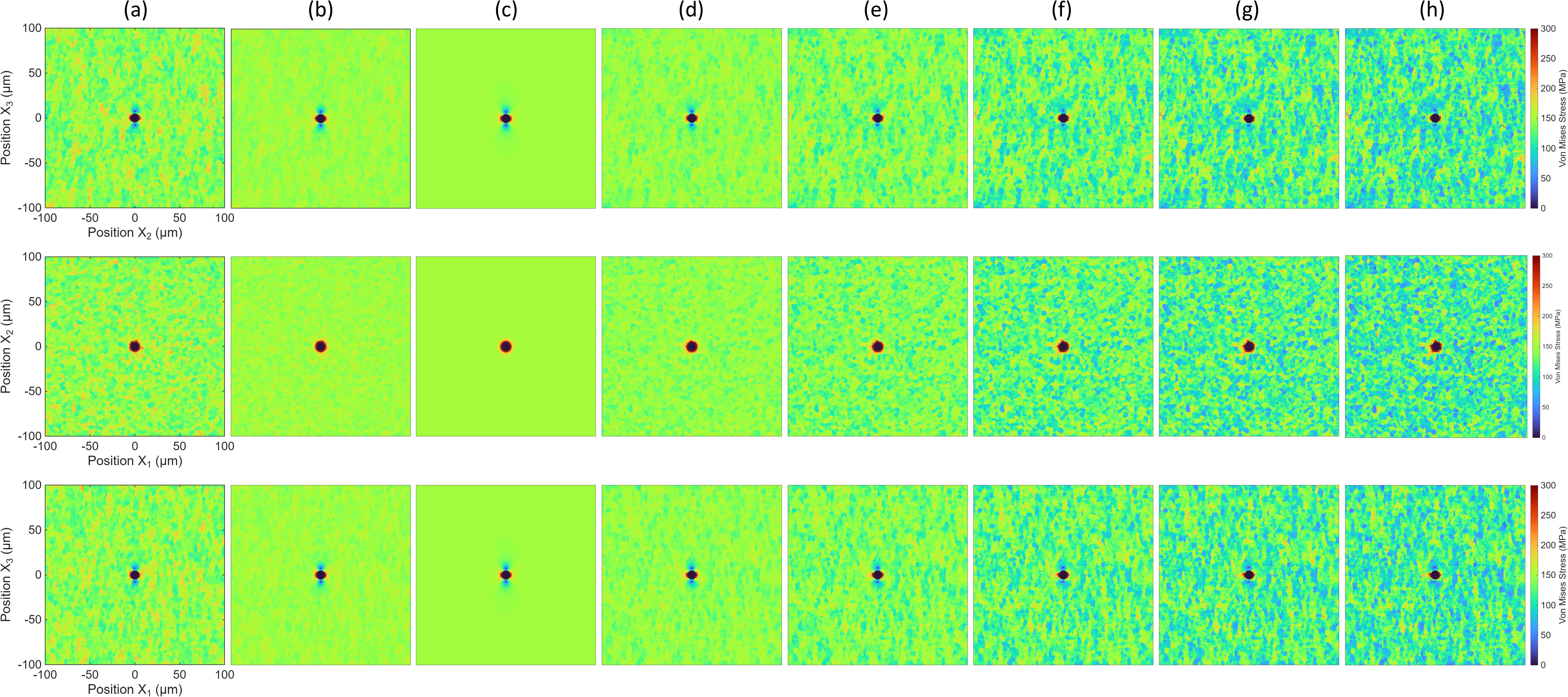}
    \caption{Evolution of the von Mises stress field around a spherical void under uniaxial tension (0.1\% strain along $X_3$-direction). The material parameters were normalized to constant average moduli ($K, G$) to isolate the effect of anisotropy. (a) $A = 0.50$, (b) $A = 0.75$, (c) $A = 1.0$ (isotropic baseline), (d) $A = 1.50$, (e) $A = 2.0$, (f) $A = 3.0$, (g) $A = 4.0$, (h) $A = 5.0$. Note the transition from a smooth continuum stress field at $A=1.0$ to highly heterogeneous stress concentrations at grain boundaries as $A$ increases. The rows correspond to central cross-sections in three orthogonal planes: top row shows the $X_2-X_3$ plane at the center of $X_1$; the middle row shows the $X_1-X_2$ plane at the center of $X_3$; and the bottom row shows the $X_1-X_3$ plane at the center of $X_2$.}
    \label{fig:8}
\end{figure}

The evolution of the von Mises stress field with increasing anisotropy is visualized in Figure~\ref{fig:8}. At the isotropic baseline ($A=1.0$, Case 3), the stress field exhibits the classic smooth pattern expected from continuum mechanics. As $A$ deviates from unity, the stress field becomes increasingly heterogeneous. 

To quantify the impact of this heterogeneity, the stress concentration factor ($K_t$) was calculated for each case by normalizing the maximum von Mises stress from the simulations to the nominal applied stress ($\sigma_{nom}$). In the case of a spherical void subjected to uniaxial strain, the stress concentration is driven by the far-field deviatoric stress component. The nominal stress was determined using the following equation: $\sigma_{nom} = \frac{8}{3} G \epsilon_{33}$~\cite{goodier1933concentration}. By substituting the baseline shear modulus of ($75.08$ GPa) and the applied strain of ($0.1\%$), the nominal stress is approximately $200.21$ MPa. 

The maximum von Mises stress ($\sigma_{max}$) utilized for the calculation of $K_t$ was derived from the full-field solution. In voxel-based finite element meshes, staircase artifacts at curved interfaces can lead to the formation of artificial stress singularities. To accurately determine maximum stress, a 3D Gaussian smoothing operation is applied to the raw stress field. The kernel's standard deviation ($s_{kernel}$) of the smoothing operation controls the degree of smoothing. In this study, $s = 0.4$ voxel unit is chosen to balance noise suppression and the preservation of genuine stress concentrations. A separate discussion of the 3D Gaussian smoothing technique applied to the von Mises stress field is provided in~\ref{app3}.

\begin{figure}[htbp]
    \centering
    \includegraphics[width=0.7\textwidth]{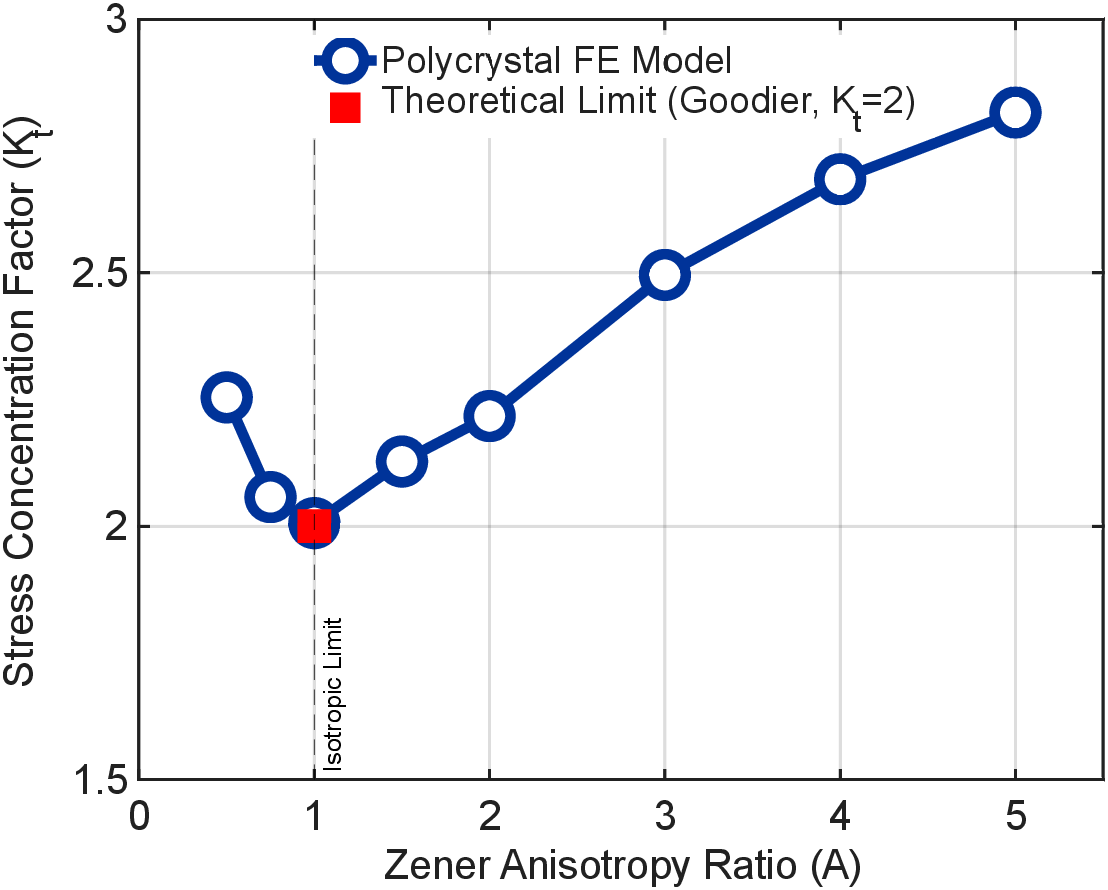}
    \caption{Sensitivity of the stress concentration factor ($K_t$) to the Zener anisotropy ratio ($A$) for a spherical void in a polycrystalline medium. The red marker indicates the theoretical isotropic limit (Goodier's solution) at $A=1.0$. Deviations from unity in either direction ($A<1$ or $A>1$) increase the local stress concentration due to microstructural stiffness mismatch.}
    \label{fig:9}
\end{figure}

The results are summarized in Figure~\ref{fig:9}, which plots the stress concentration factor ($K_t=\sigma_{max} / \sigma_{nom}$) against the Zener ratio. The results exhibit a distinct V-shaped behavior, with a minimum at $A=1.0$. At the isotropic baseline ($A=1.0$), the maximum stress was 401.63 MPa, generating a $K_t$ of 2.006, which is in close agreement with the theoretical value of 2.0 derived from Goodier's solution for a spherical void subjected to uniaxial strain~\cite{goodier1933concentration,mura2013micromechanics}. 

The stress concentration factor increases monotonically as anisotropy becomes more pronounced, regardless of whether the material deviates towards $A < 1$ or $A > 1$.
For the cases where $A < 1$, the maximum von Mises stress increases to 411.91 MPa ($K_t$ = 2.057) at $A=0.75$ and further to 451.27 MPa ($K_t$ = 2.254) at $A=0.50$. Similarly, for the cases where $A > 1$, the maximum stress increases to 425.94 MPa ($K_t$ = 2.127) at $A=1.5$, 443.90 MPa ($K_t$ = 2.217) at $A=2.0$, and increases sharply to 499.62 MPa ($K_t$ = 2.495) at $A=3.0$. At higher levels of anisotropy, the trend continues, with stress reaching 537.41 MPa ($K_t$ = 2.684) at $A=4.0$ and peaking at 563.78 MPa ($K_t$ = 2.816) for $A=5.0$. This nearly 40\% increase relative to the isotropic case occurs because the applied load is preferentially transmitted through grains oriented along the stiff $\langle 111 \rangle$ directions. However, in the cases where $A < 1$, stress hotspots tend to rotate in order to align with the stiff directions $\langle 100 \rangle$. The observation indicates that crystallographic anisotropy intensifies local stress concentrations beyond any deviation from isotropy, regardless of whether the anisotropy favors $\langle 100 \rangle$ or $\langle 111 \rangle$ orientations.

\subsection{Effects of microstructure and defect type on stress mapping in LPBF SS316L}
\label{sec3:3}
\begin{figure}[htbp]
  \centering
  \includegraphics[width=1.0\textwidth]{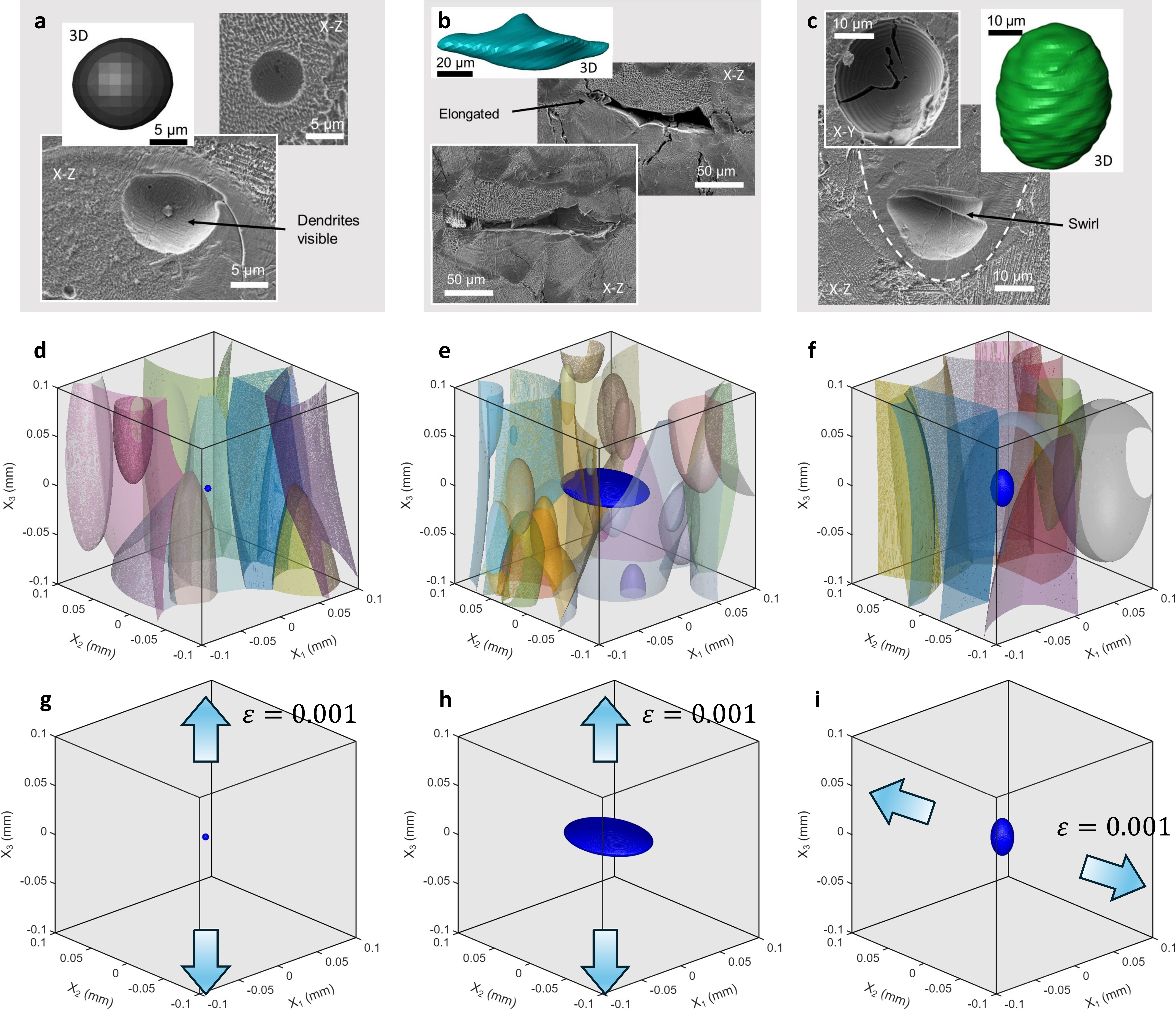}
  \caption{Comparison of defect morphologies. The top row displays experimental observations reproduced from Boig~\cite{boig2019application}, while the bottom row shows the corresponding simulation models. The columns illustrate (from left to right): Lack of Fusion (LoF) voids, modeled as flattened \textit{oblate spheroids} to mimic unbonded layers; Gas voids, modeled as \textit{spheres} typical of trapped gas; and Keyhole voids, modeled as elongated \textit{prolate spheroids} to represent deep laser penetration.}
  \label{fig:10}
\end{figure}

In order to accurately capture the mechanical response of additively manufactured material, the simulation incorporates three distinct defect morphologies experimentally identified, which are classified according to the laser energy density regimes governing their formation. Figures~\ref{fig:10}(a--c) compile representative scanning electron microscopy (SEM) micrographs and 3D XCT reconstructions of gas pores, lack of fusion (LoF) voids, and keyhole defects of nickel-base superalloy IN713C that are produced during the LPBF process~\cite{boig2019application}. Gas porosity can occur across the all range of input energy levels, originating from trapped gas within the feedstock powder. Although these tiny, spherical pores ($<50$ $\mu m$) are less common at elevated energy levels, they remain a consistent background defect type~\cite{du2023pore}. LoF voids occur at low energy levels, where the energy density is insufficient to produce complete melt-pool overlap. These voids typically exhibit an elongated shape, ranging from 100 to 200 $\mu m$, oriented parallel to the build layers, and contain unbonded powder with sharp features that serve as significant stress concentrators~\cite{bayat2019multiphysics}. Keyhole voids are indicative of the high-energy regime, where excessive energy input causes deep laser penetration and turbulent flow at the bottom of the melt pool. These defects typically present as smooth, spherical, or oval voids with diameters varying from 20 to 200 $\mu m$~\cite{guo2023understanding} extending vertically across layers. 

Figure~\ref{fig:10} illustrates the multi-scale modeling workflow used to evaluate these defects. The geometric reconstruction in Figures~\ref{fig:10}(d--f) utilizes a volume-conserving algorithm to translate experimental scan data into idealized 3D shapes. Gas voids are modeled as ideal spheres where the radius $r$ is half the equivalent spherical diameter ($r=d_{eq}/2$). LoF voids are modeled as flattened triaxial ellipsoids to represent inter-layer delamination and their elongated footprint along the scan track. The in-plane major semi-axis $a$ is defined as half the maximum Feret diameter $D_{Feret\_max}/2$, which is the length along the scan vector. To capture the fact that LoF pores are rarely circular in-plane, the in-plane minor semi-axis $b$ is defined as $b=a\alpha_{elongation}$, where the elongation (width-to-length) ratio $\alpha_{elongation}$ was sampled uniformly in the range $\alpha_{elongation} \in [0.5, 0.9]$. The through-thickness semi-axis along the build direction $c$ obtained from the measured pore volume $V_{pore}$ by enforcing volume consistency,  $c = V_{pore} / (\frac{4}{3}\pi a b)$. The defect orientation is applied by using rotation matrices ($R_z$ for yaw and $R_y$ for tilt) to align each void with the observed scan track direction and local surface roughness.

Keyhole voids were reconstructed as prolate spheroids formulation ($c > a=b$) to approximate the deep, narrow melt channel characteristic of keyhole-mode melting. The axial semi-axis $c$ along the build direction is set to half of the maximum Feret diameter, assuming the defect’s longest dimension aligns with the melt penetration depth. The lateral semi-axes $a$ and $b$ are obtained from the measured pore volume $V_{pore}$ by enforcing volume consistency, $a = b = \sqrt{V_{pore} / (\frac{4}{3}\pi c)}$. To ensure physical fidelity, the algorithm included a safety mechanism: if the derived width $a$ exceeded the depth $c$, implying the defect was not a deep keyhole but likely a misclassified gas pore, the model automatically reverted the geometry to an equivalent sphere ($r_{eq} = (\frac{3 V_{pore}}{4 \pi})^{1/3}$).

\begin{figure}[htbp]
  \centering
  \includegraphics[width=0.9\textwidth]{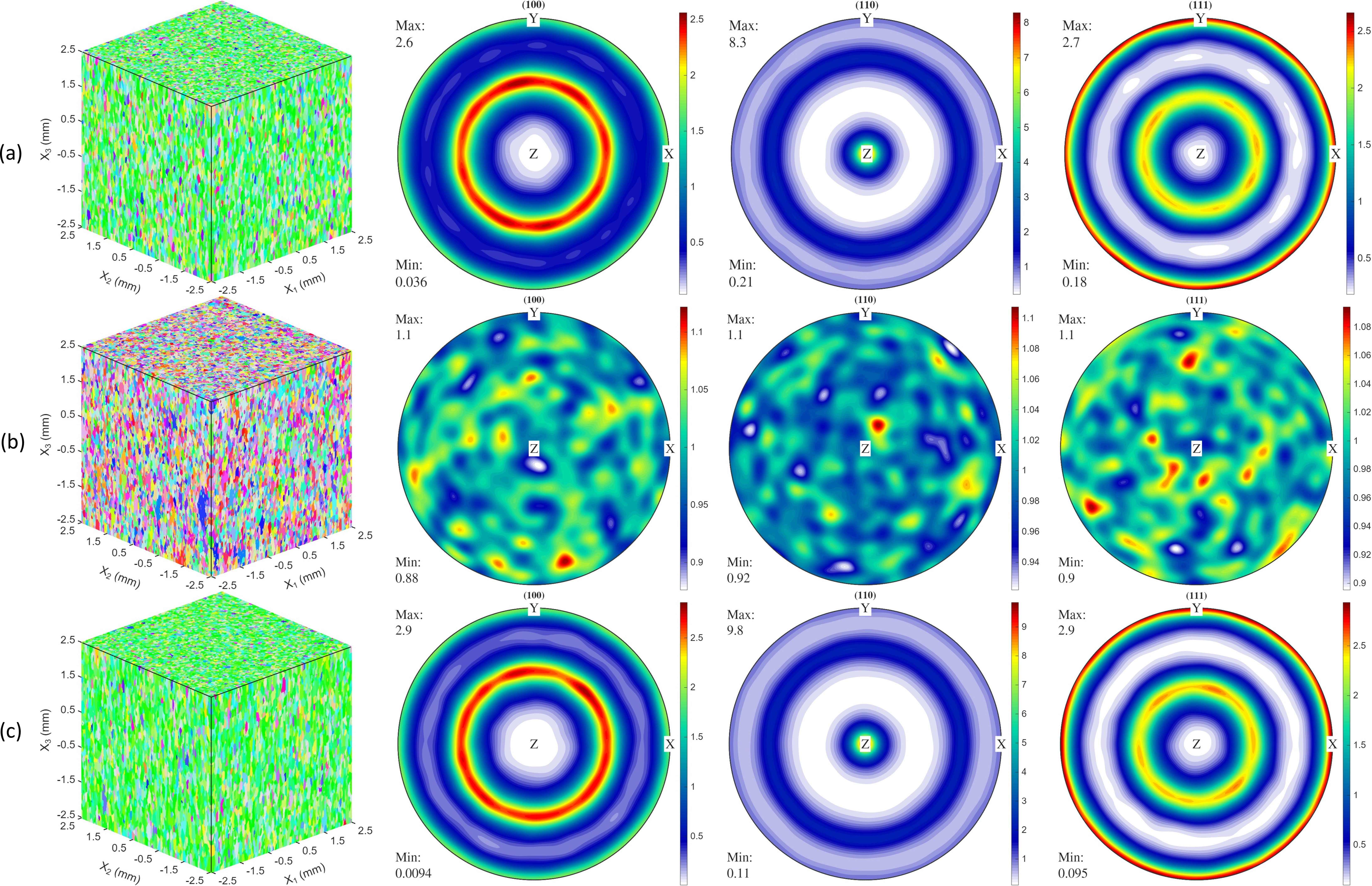}
  \caption{Synthetic 3D microstructure and texture validation for different energy regimes. (a) Gas void: strong [101] texture with columnar grains, (b) LoF void: random texture with smaller, less elongated grains, (c) KH void: strong [101] texture with large columnar grains.}
  \label{fig:11}
\end{figure}

These idealized defects are embedded at the center of $200 \times 200 \times 200$ $\mu m^3$ polycrystalline RVEs extracted from the larger domains shown in Figure~\ref{fig:11}. To ensure physical fidelity, the microstructure for each case was synthesized to match the specific grain statistics and texture observed in the corresponding energy regime. For the gas and keyhole cases shown in Figure~\ref{fig:11}(a) and (c), which correspond to optimal and high-energy processing, the microstructure exhibits strong columnar grain growth aligned with the thermal gradient. These regimes feature large grains with average sizes of $55 \mu m$ and $56.5 \mu m$, and high aspect ratios of 3.18 and 3.25, respectively. They exhibit a sharp $[101]$ fiber texture along the build direction with spread $\approx 12.5^{\circ}-15^{\circ}$, evidenced by the predominant green hue in the Inverse Pole Figure (IPF-Z) maps. The distinct intensity rings are identified in the pole figures. In contrast, the LoF regime depicted in Figure~\ref{fig:11}(b) is characterized by insufficient thermal input to sustain competitive columnar growth. Consequently, it exhibits smaller grains with an average size of $45 \mu m$, a lower aspect ratio of 2.65, and a random texture, visualized as a rainbow pattern in the IPF maps and as evenly scattered points in the pole figure.

To ensure that the stress gradients at the sharp defect tips are resolved with high fidelity, the computational domain for each pore case was discretized into a high-resolution voxel grid of $800 \times 800 \times 800$ voxels. The elastic constants $C_{ij}$ utilized for the cubic stainless steel 316L phase are summarized in Table~\ref{tab:5}. 

\begin{table}[htbp]
    \centering
    \caption{Elastic stiffness components for single-crystal stainless steel 316L~\cite{ledbetter1981predicted,charmi2021mechanical}.}
    \label{tab:5}
    \begin{tabular}{ccc}
        \toprule
        $C_{11}$ (GPa) & $C_{12}$ (GPa) & $C_{44}$ (GPa) \\
        \midrule
        206 & 133 & 119 \\
        \bottomrule
    \end{tabular}
\end{table}

\begin{figure}[htbp]
  \centering
  \includegraphics[width=0.7\textwidth]{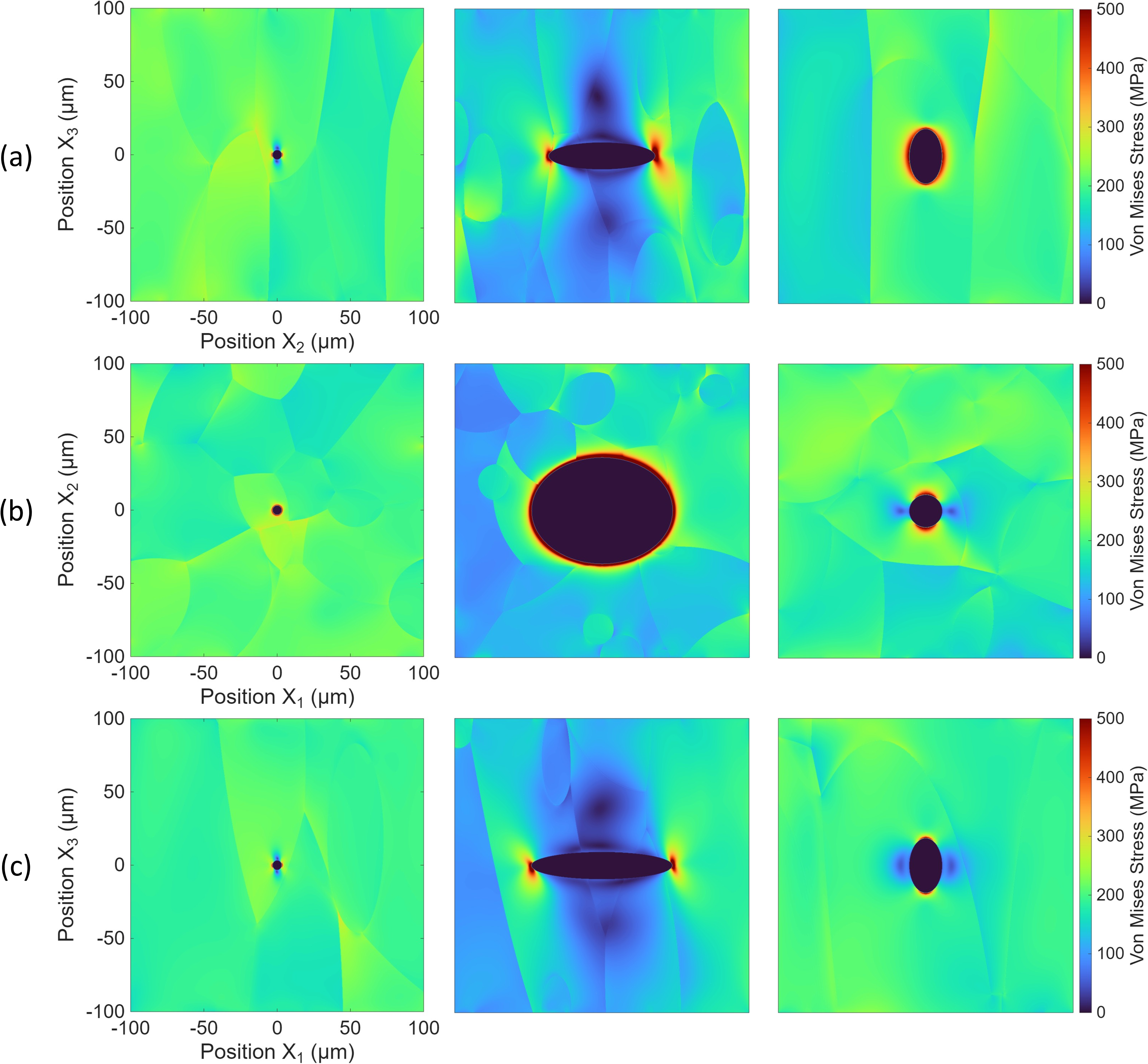}
  \caption{von Mises stress mapping (MPa) computed by the PCG solver for three defect types: Gas pore (left column), Lack of Fusion (middle column), and Keyhole void (right column). The rows demonstrate different cross-sectional views through the void center: (a) the $X_2$--$X_3$ plane (top row); (b) the $X_1$--$X_2$ plane (middle row); and (c) the $X_1$--$X_3$ plane (bottom row).}
  \label{fig:12}
\end{figure}

\begin{figure}[htbp]
  \centering
  \includegraphics[width=1.0\textwidth]{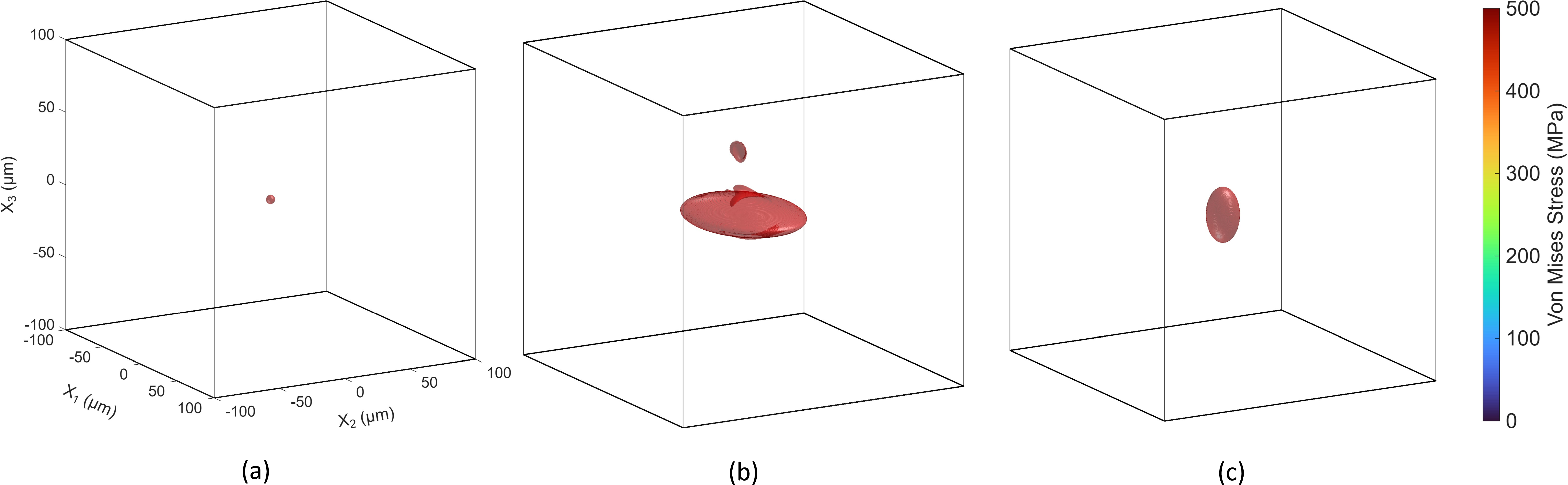}
  \caption{von Mises 3D stress mapping (MPa) (top 1\% high stress) computed by the PCG solver for three defect types: (a) Gas pore, (b) Lack of Fusion, and (c) Keyhole void.}
  \label{fig:13}
\end{figure}

Figures~\ref{fig:12} and~\ref{fig:13} show the von Mises stress fields for the three defect types. Figure~\ref{fig:12} presents 2D cross-sections, while Figure~\ref{fig:13} shows 3D iso-surfaces highlighting the top 1\% of peak stress. The high mesh resolution ($800^3$ voxels) captures how defect shape and loading direction interact with the surrounding polycrystalline microstructure. The loading conditions applied to these RVEs in Figure~\ref{fig:10}(g--i) are selected to simulate the worst-case scenario for crack initiation specific to each morphology: arbitrary uniaxial loading for gas pores, $X_3$-direction (Mode I) loading for planar LoF voids, and transverse loading $X_1$ for cylindrical Keyholes.

The numerical results establish a clear hierarchy of defect severity. The gas pore in Figure~\ref{fig:13}(a) exhibits the least intense response, producing a nearly isotropic stress distribution with a modest stress concentration limited to small volumes affected by grain boundary stiffness mismatches. The keyhole defect in Figure~\ref{fig:13}(c) shows intermediate severity. However, transverse loading causes stress trajectories to crowd around the curved boundary, analogous to a hole in a plate, leading to a high stress concentration factor. The LoF void shown in Figure~\ref{fig:13}(b) demonstrates the highest severity. Under  $X_3$-direction loading, the flattened geometry in the $X_1-X_2$ plane generates a severe stress singularity at the sharp crack tips (Mode I opening)~\cite{beretta2017comparison, alfred2025high}, resulting in a continuous ring of yield-level stress that far exceeds the concentrations observed in the other cases.

\section{Discussions}
\label{sec4}
The findings detailed in Section~\ref{sec3:1} demonstrate that the PCG-based \textit{ELAS3D-Xtal} solver accurately reproduces the analytical solution for an isotropic solid that includes a spherical void, achieving both high precision and significant computational efficiency. In addition to validating stress magnitudes, the primary stress trajectories illustrated in Figure~\ref{fig:app2} provide a structural perspective on load transfer through the matrix and elucidate the physical origins of stress concentrations around porosity. 

In the side-view planes as shown in Figures~\ref{fig:app2}(a) and (c), the trajectories corresponding to the maximum principal stress curve around the void indicate that the axial load is redirected through the adjacent matrix. These trajectories are densely concentrated near the void equator, aligning with the region of maximum von Mises stress. This suggests that the associated stress concentration is directly responsible for the crack initiation. Near the void poles, the trajectories spread outward, consistent with reduced local stresses in these regions.

The patterns of orthogonal circumferential and radial trajectories appear to be a fundamental topological response of continuous materials to geometric singularities, extending beyond simple voids. Similar topological structures have been reported for indentation in polycrystalline copper~\cite{voyiadjis2021grain} and single crystals~\cite{saito2011wedge}. In those studies, the field near the indenter contact point is organized into two orthogonal families: a radial set associated with the minimum principal stretch and a circumferential set associated with the maximum principal stretch. Kysar et al.~\cite{kysar2010experimental} argued that the close relationship between indentation and fracture reflects the rapid spatial variations in the deformation field that develop near singularities. This perspective suggests that such trajectory formations can serve as indicators of stress concentrations, whether caused by open-volume defects, as in Section~\ref{sec3:1}, or by rigid contact singularities. Establishing this topological reference for the isotropic Eshelby problem is therefore important for interpreting the complex, distorted stress fields that emerge under elastic anisotropy.

The sensitivity analysis in Section~\ref{sec3:2} shows that the V-shaped trend in the stress concentration factor ($K_t$) demonstrates that any deviation from elastic isotropy ($A=1$) increases local stresses. For materials with $A > 1.0$ (typical of FCC metals such as SS316L, Copper), the $\langle 111 \rangle$ crystal direction becomes significantly stiffer than the $\langle 100 \rangle$ direction. The load is preferentially carried through grains where the stiff $\langle 111 \rangle$ axis aligns with the loading direction. This creates sharp stress gradients at grain boundaries that superimpose onto the geometric stress field of the void. 
In contrast, for materials with $A < 1.0$ (e.g., Niobium), the stiffness trend inverts: $C_{11}$ becomes high relative to $C_{44}$, making the $\langle 100 \rangle$ direction the primary load carrier. As a result, stress hotspots appear in spatially distinct locations, rotated by approximately $45^{\circ}$ relative to the lattice compared to the $A > 1$ cases~\cite{chiang2007stress}. This observation suggests that in real-world additive manufacturing situations, the critical point of failure is a combined function of both pore geometry and the orientation of local grains~\cite{qiu2024new}.

Building on idealized anisotropy results, simulations of realistic LPBF defects in Section~\ref{sec3:3} show that defect geometry often has a stronger influence on failure than crystallographic texture. Although the anisotropy of SS316L ($A > 1$) promotes load localization along the stiff $\langle 111 \rangle$ directions, the LoF defect dominates the response, producing a stress singularity that far exceeds the local fluctuations caused by grain orientation~\cite{riemer2014fatigue}. The flattened LoF morphology behaves as a pre-existing crack (Mode I), leading to a stress intensity that is more critical than the geometric concentration of a keyhole defect or the comparatively benign field around a spherical gas pore~\cite{beretta2017comparison}. While crystallographic texture can influence the preferred load-transfer pathways through the matrix, structural durability of the printed component is governed by the presence of these planar LoF defects above all other microstructural features~\cite{sanaei2021defects}.

\section{Conclusions}
\label{sec5}
In this work, we presented \texttt{ELAS3D-Xtal}, a high-performance computational framework that bridges experimental microstructure characterization and predictive elastic mechanics for additively manufactured materials. By modernizing the legacy NIST ELAS3D architecture with OpenMP parallelization, improved iterative solution strategies, and HDF5 I/O, we address key scalability and convergence limitations that previously hindered large-scale simulations of high-contrast heterogeneous media.

The advances reported here enable three primary capabilities. First, a matrix-free PCG solver with a point-block Jacobi preconditioner improves convergence robustness in porous domains by mitigating the ill-conditioning associated with the extreme stiffness contrast between solid metal and voids. Second, a native parallel microstructure generator enables rapid construction of statistically RVEs with user-defined grain statistics, crystallographic textures, and explicit pore populations derived from XCT measurements. Third, shared-memory parallel implementation supports efficient full-field stress calculations on voxel grids containing hundreds of millions of degrees of freedom on a single multicore workstation.

Validation against the analytical Eshelby inclusion solution demonstrates strong agreement, with remaining discrepancies primarily confined to the voxelized inclusion boundary, and mesh studies confirm stable behavior across resolutions. Application studies further indicate that elastic anisotropy and defect morphology, such as gas, lack-of-fusion, and keyhole pores, significantly influence local stress concentrations in LPBF SS316L. This highlights the need to address coupled defect--microstructure effects when generating fatigue-relevant stress metrics.

Future work will extend this elastic framework into a multi-fidelity microstructure-based life prediction workflow that links process-induced defects directly to fatigue performance. This workflow will proceed in three stages: (i) using \texttt{ELAS3D-Xtal} for rapid elastic screening to identify critical crack-initiation sites and compute crack-driving-force metrics for small cracks emanating from pore surfaces; (ii) analyzing selected critical sub-volumes with crystal plasticity solvers to quantify plastic energy storage and fatigue indicator parameters for crack nucleation, enabling calibration against experimentally observed defect--fatigue correlations; and (iii) exploring graph-based approaches~\cite{srivastava2021graph} to predict three-dimensional, microstructure-sensitive crack propagation paths through heterogeneous grain structures. This integrated approach establishes a direct computational pathway from as-built defect populations to component-scale fatigue reliability.

\section*{Declaration of competing interest}
The authors declare that they have no known competing financial interests or personal relationships that could have appeared to influence the work reported in this paper.

\section*{Acknowledgements}
This work was supported by the Defense Advanced Research Projects Agency (DARPA) SURGE program under Cooperative Agreement No. HR0011-25-2-0009, `Predictive Real-time Intelligence for Metallic Endurance (PRIME)'.

\appendix
\section{Mesh sensitivity analysis}
\label{app1}

The agreement between the analytical field shown in Figure \ref{fig:app1_1}(b) and the numerical field in Figure \ref{fig:app1_1}(c) across all resolutions demonstrates the convergence and stability of the solver. The results shown in Figure \ref{fig:app1_2} confirm that the PCG solver accurately reproduces the analytical stress fields across different cross-sectional orientations and mesh resolutions.

\setcounter{figure}{0} \renewcommand{\thefigure}{A.\arabic{figure}}
\begin{figure}[ht]
  \centering
  \includegraphics[width=1.0\textwidth]{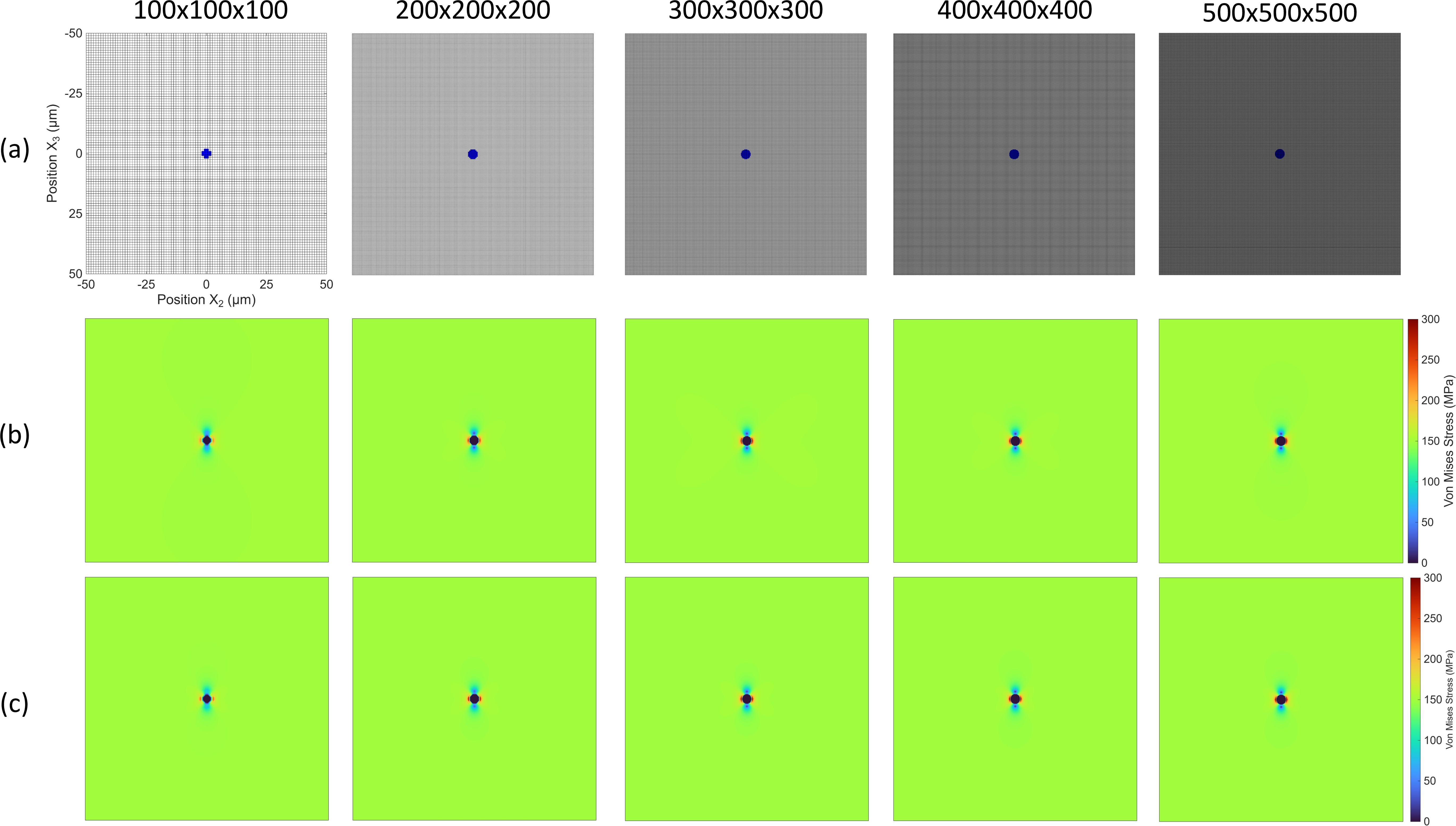}
  \caption{Mesh independence verification for the 3D Eshelby inclusion problem described in Section~\ref{sec3:1}, comparing analytical and numerical von Mises stress distributions (MPa). The five columns correspond to increasing mesh densities of $100^3$, $200^3$, $300^3$, $400^3$, and $500^3$ voxels, respectively.
  \textbf{(a)} Phase maps of the $X_2-X_3$ cross-section at the central $X_1$-slice, showing the grid resolution;
  \textbf{(b)} The analytical Eshelby solution; and
  \textbf{(c)} The numerical solution computed using the PCG solver.}
  \label{fig:app1_1}
\end{figure}

\setcounter{figure}{1} \renewcommand{\thefigure}{A.\arabic{figure}}
\begin{figure}[ht]
  \centering
  \includegraphics[width=1.0\textwidth]{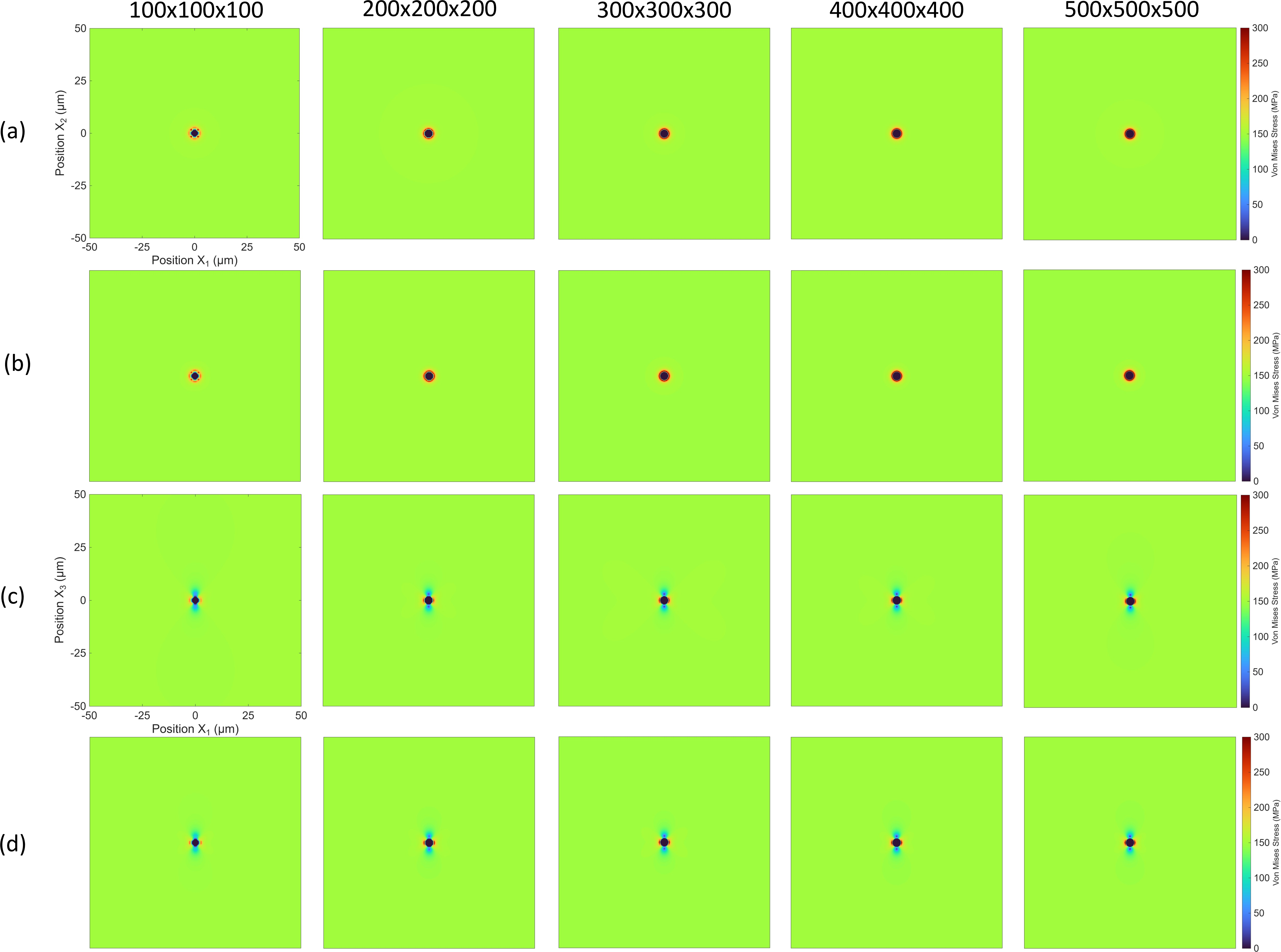}
  \caption{Extended mesh independence verification comparing analytical and numerical von Mises stress distributions (MPa). As shown in Figure~\ref{fig:app1_1}, the five columns represent increasing mesh densities, ranging from $100^3$ to $500^3$.
  \textbf{(a--b) $X_1-X_2$} Plane: Comparison at the central $X_3$-slice between (a) the analytical Eshelby solution and (b) the numerical PCG solution.
  \textbf{(c--d) $X_1-X_3$} Plane: Comparison at the central $X_2$-slice between (c) the analytical Eshelby solution and (d) the numerical PCG solution.}
  \label{fig:app1_2}
\end{figure}

\section{Analysis of principal stress trajectories (isostatics)}
\label{app2}

Figure~\ref{fig:app2} describes the principal stress trajectories across three orthogonal planes subjected to uniaxial loading along $X_3$. In Figure~\ref{fig:app2}(a), the $X_2$–-$X_3$ plane (side view; mid-$X_1$ slice) illustrates vertical load paths that divert around the void, along with transverse confinement trajectories located near the matrix–void interface. Figure~\ref{fig:app2}(b) presents the $X_1$–-$X_2$ plane (top view; mid-$X_3$ slice), which reveals an orthogonal network of circumferential and radial trajectories within the plane normal to the loading axis. Figure~\ref{fig:app2}(c) shows the $X_1$-–$X_3$ plane (side view; mid-$X_2$ slice), which validates the expected symmetry in the patterns of load diversion and stress concentration. The background colormap illustrates the magnitude of von Mises stress (MPa).

The stress trajectories are computed by first solving the eigenvalue problem of the stress tensor field at each mesh point to obtain principal directions, then integrating these vector fields to generate 2D projected trajectories. For each selected plane, the relevant in-plane components are extracted from the full 3D stress tensor to form the corresponding 2D stress tensor, $\mathbf{\sigma}_{2D} = \begin{bmatrix} \sigma_{11} & \tau_{13} \\ \tau_{13} & \sigma_{33} \end{bmatrix}$, where $\tau_{13}$ is the shear stress. The principal directions are obtained by diagonalizing $\mathbf{\sigma}_{2D}$. The orientation angle $\theta$ of the principal axes relative to the horizontal axis is computed via eigen-decomposition as: $\theta(X_1,X_3) = \frac{1}{2} \arctan\left( \frac{2\tau_{13}}{\sigma_{11} - \sigma_{33}} \right)$. At each mesh point, this yields two orthogonal unit vectors: $\mathbf{v}_a = \begin{bmatrix} \cos\theta \\ \sin\theta \end{bmatrix}$ and $\mathbf{v}_b = \begin{bmatrix} -\sin\theta \\ \cos\theta \end{bmatrix}$. These vectors are sorted to ensure globally coherent trajectory families, and the resulting continuous vector fields are integrated to construct the trajectories. The path $\mathbf{r}(s)$ of a trajectory satisfies $\frac{d\mathbf{r}}{ds} = \mathbf{v}(\mathbf{r})$, where $s$ is the arc length. This differential equation is solved numerically via a Runge-Kutta method to trace the curves shown in Figure~\ref{fig:app2}.

\setcounter{figure}{2} \renewcommand{\thefigure}{B}
\begin{figure}[ht]
  \centering
  \includegraphics[width=1.0\textwidth]{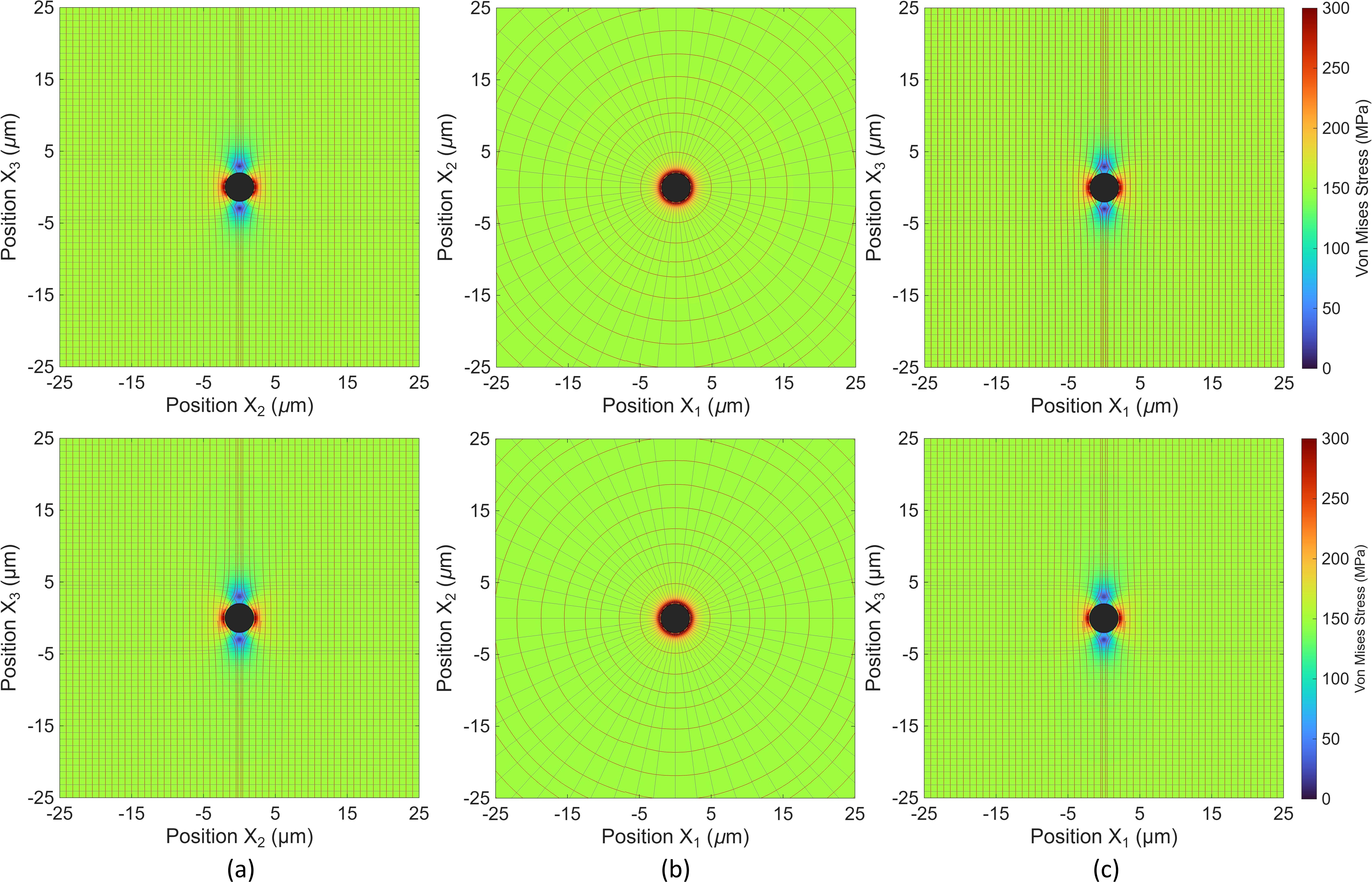}
  \caption{Visualization of principal stress trajectories (isostatics) superimposed on von Mises stress distributions from the (a) analytical and (b) numerical PCG solutions around a central void. Shown are the X2–X3 cross-section at the mid-X1 slice (left column), the X1–X2 cross-section at the mid-X3 slice (middle column), and (c) the X1–X3 cross-section at the mid-X2 slice (right column). Visualization of principal stress trajectories (isostatics) superimposed on von Mises stress distributions from the analytical (top row) and numerical PCG (bottom row) solutions around a central void. (a) X2–X3 cross-section at the mid-X1 slice; (b) X1–X2 cross-section at the mid-X3 slice; (c) X1–X3 cross-section at the mid-X2 slice.}
  \label{fig:app2}
\end{figure}

\section{Post-processing and maximum stress extraction}
\label{app3}
To mitigate the effects of voxelization artifacts inherent to the regular-grid discretization of curved interfaces, a consistent smoothing algorithm is applied to the raw simulation output prior to extracting maximum stress values. The raw von Mises stress field, defined on a grid of $N_x \times N_y \times N_z$ voxels, is processed using a 3D Gaussian filter (MATLAB function \textit{imgaussfilt3}). To prevent zero-stress values within pores from biasing the stress in the surrounding material, a normalized convolution approach is implemented. First, a labeled phase mask $M(x,y,z)$ was defined, where $M=0$ denotes the void and $M=g \in \{1,\ldots,N_g\}$ denotes voxels belonging to grain $g$ (with $N_g$ the total number of grains). 
The raw stress field $\sigma_{raw}$ is masked, with the pore values set to zero. A Gaussian kernel with a standard deviation of $s_{kernel} = 0.4$ voxels is applied separately to both the masked stress field and the mask itself:\begin{equation}\sigma_{smooth} = \frac{G_{s} * (\sigma_{raw} \cdot M)}{G_{s} * M}\end{equation}where $G_{s}$ denotes the Gaussian kernel parameterized by width $s_{kernel}$, and $*$ denotes the convolution operation. This approach guarantees that the smoothing at the pore-solid interface is normalized by the contributions from neighboring solid voxels. 
After the smoothing process, local maxima are detected using a 26-connected neighborhood search (via MATLAB function \textit{imregionalmax}). The maximum von Mises stress ($\sigma_{max}$) described in Section~\ref{sec3:2} corresponds to the maximum value found among these hotspots within the solid phases. This methodology ensures that the identified stress concentrations reflect physical hotspots arising from microstructural heterogeneity rather than numerical singularities at voxel corners.

\end{document}